\documentclass{emulateapj}
\usepackage{apjfonts}
\usepackage{color}

\def\asca{{\itshape ASCA\/}}

\def\chandra{{\itshape Chandra\/}}

\def\hst{{\itshape HST\/}}

\def\rxte{{\itshape RXTE\/}}

\def\spitzer{{\itshape Spitzer\/}}

\def\galex{{\itshape GALEX\/}}

\def\xmm{{\itshape XMM-Newton\/}}
\def\nustar{{\itshape NuSTAR\/}}

\def\xray{\hbox{X-ray}}

\def\etal{{et\,al.}}

\def\ltsima{$\; \buildrel < \over \sim \;$}
\def\simlt{\lower.5ex\hbox{\ltsima}}
\def\gtsima{$\; \buildrel > \over \sim \;$}
\def\simgt{\lower.5ex\hbox{\gtsima}}
\def\kms{\ifmmode{~{\rm km~s^{-1}}}\else{~km s$^{-1}$}\fi}
\def\lsim{\lower0.3em\hbox{$\,\buildrel <\over\sim\,$}}
\def\gsim{\lower0.3em\hbox{$\,\buildrel >\over\sim\,$}}

\def\msol{$M_\odot$}
\def\h2{H$_2$}
\def\flux{erg~cm$^{-2}$~s$^{-1}$}
\def\lum{erg~s$^{-1}$}

\def\sfr{$M_{\odot}$~yr$^{-1}$}

\def\aap{A\&A}
\def\apj{ApJ}
\def\apjl{ApJL}
\def\apjs{ApJS}
\def\aj{AJ}
\def\mnras{MNRAS}
\def\araa{ARA\&A}
\def\pasj{PASJ}
\def\nat{Nature}

\begin{document}

\shortauthors{LEHMER ET AL.}
\shorttitle{\nustar\ and \chandra\ View of NGC~3256 and NGC~3310}

%
\title{The 0.3--30~keV Spectra of Powerful Starburst Galaxies: {\itshape NUSTAR} and {\itshape CHANDRA} Observations of NGC~3256 and NGC~3310}
%

\author{
B.~D.~Lehmer,\altaffilmark{1,2}
J.~B.~Tyler,\altaffilmark{2,3}
A.~E.~Hornschemeier,\altaffilmark{2}
D.~R.~Wik,\altaffilmark{1,2}
M.~Yukita,\altaffilmark{1,2}
V.~Antoniou,\altaffilmark{4}
S.~Boggs,\altaffilmark{5}
F.~E.~Christensen,\altaffilmark{6}
W.~W.~Craig,\altaffilmark{5,7}
C.~J.~Hailey,\altaffilmark{8}
F.~A.~Harrison,\altaffilmark{9}
T.~J.~Maccarone,\altaffilmark{10}
A.~Ptak,\altaffilmark{2}
D.~Stern,\altaffilmark{11}
A.~Zezas,\altaffilmark{12,13,4}
\& 
W.~W.~Zhang\altaffilmark{2}
}

\altaffiltext{1}{The Johns Hopkins University, Homewood Campus, Baltimore, MD 21218, USA}
\altaffiltext{2}{NASA Goddard Space Flight Center, Code 662, Greenbelt, MD 20771, USA} 
\altaffiltext{3}{Institute for Astrophysics and Computational Sciences,
Department of Physics, The Catholic University of America, Washington, DC
20064, USA}
\altaffiltext{4}{Harvard-Smithsonian Center for Astrophysics, 60 Garden Street, Cambridge, MA 02138, USA}
\altaffiltext{5}{Space Sciences Laboratory, University of California, Berkeley, CA 94720, USA}
\altaffiltext{6}{DTU Space - National Space Institute, Technical University of Denmark, Elektrovej 327, 2800 Lyngby, Denmark}
\altaffiltext{7}{Lawrence Livermore National Laboratory, Livermore, CA 94720, USA}
\altaffiltext{8}{Columbia Astrophysics Laboratory, Columbia University, New York, NY 10027, USA}
\altaffiltext{9}{Caltech Division of Physics, Mathematics and Astronomy, Pasadena, USA}
\altaffiltext{10}{Department of Physics, Texas Tech University, Lubbock, TX 79409, USA}
\altaffiltext{11}{Jet Propulsion Laboratory, California Institute of Technology, Pasadena, CA 91109, USA}
\altaffiltext{12}{Physics Department \& Institute of Theoretical \& Computational Physics, University of Crete, 71003 Heraklion, Crete, Greece} 
\altaffiltext{13}{Foundation for Research and Technology-Hellas, 71110 Heraklion, Crete, Greece}

%
\begin{abstract}
%

We present nearly simultaneous \chandra\ and \nustar\ observations of two
actively star-forming galaxies within 50~Mpc: NGC~3256 and NGC~3310.  Both
galaxies are significantly detected by both \chandra\ and \nustar, which
together provide the first-ever spectra of these two galaxies spanning
\hbox{0.3--30~keV}.  The \xray\ emission from both galaxies is spatially
resolved by \chandra; we find that hot gas dominates the $E <$~\hbox{1--3~keV}
emission while ultraluminous \xray\ sources (ULXs) provide majority
contributions to the emission at $E >$~\hbox{1--3~keV}.  The \nustar\
galaxy-wide spectra of both galaxies follow steep power-law distributions with
$\Gamma \approx 2.6$ at $E >$~\hbox{5--7}~keV.
Using
new and archival \chandra\ data, we search for signatures of heavily obscured
or low luminosity AGN.  We find that both NGC~3256 and NGC~3310 have \xray\
detected sources coincident with nuclear regions; however, the steep \nustar\
spectra of both galaxies restricts these sources to be either low luminosity
AGN ($L_{\rm 2-10~keV}/L_{\rm Edd} \simlt 10^{-5}$) or non-AGN in nature (e.g.,
ULXs or crowded \xray\ sources that reach $L_{\rm 2-10~keV} \sim 10^{40}$~\lum\
cannot be ruled out).  Combining our constraints on the \hbox{0.3--30~keV}
spectra of NGC~3256 and NGC~3310 with equivalent measurements for nearby
star-forming galaxies M83 and NGC~253, we analyze the star-formation rate (SFR)
normalized spectra of these starburst galaxies.  The spectra of all four
galaxies show sharply declining power-law slopes at energies above 3--6~keV
primarily due to ULX populations.  Our observations therefore constrain the
average spectral shape of galaxy-wide populations of luminous accreting
binaries (i.e., ULXs).  Interestingly, despite a completely different 
galaxy sample selection, emphasizing here a range of SFRs and stellar masses,
these properties are similar to those of super-Eddington accreting ULXs that
have been studied individually in a targeted \nustar\ ULX program.  We also
find that NGC~3310 exhibits a factor of \hbox{$\approx$3--10} elevation of
\xray\ emission over the other star-forming galaxies due to a corresponding
overabundance of ULXs.  We argue that the excess of ULXs in NGC~3310 is most
likely explained by the relatively low metallicity of the young stellar
population in this galaxy, a property that is expected to produce an excess of
luminous \xray\ binaries for a given SFR.

%
\end{abstract}
%


%
\section{Introduction}
%

%
%
\begin{figure*}
\figurenum{1}
\centerline{
\includegraphics[width=9.2cm]{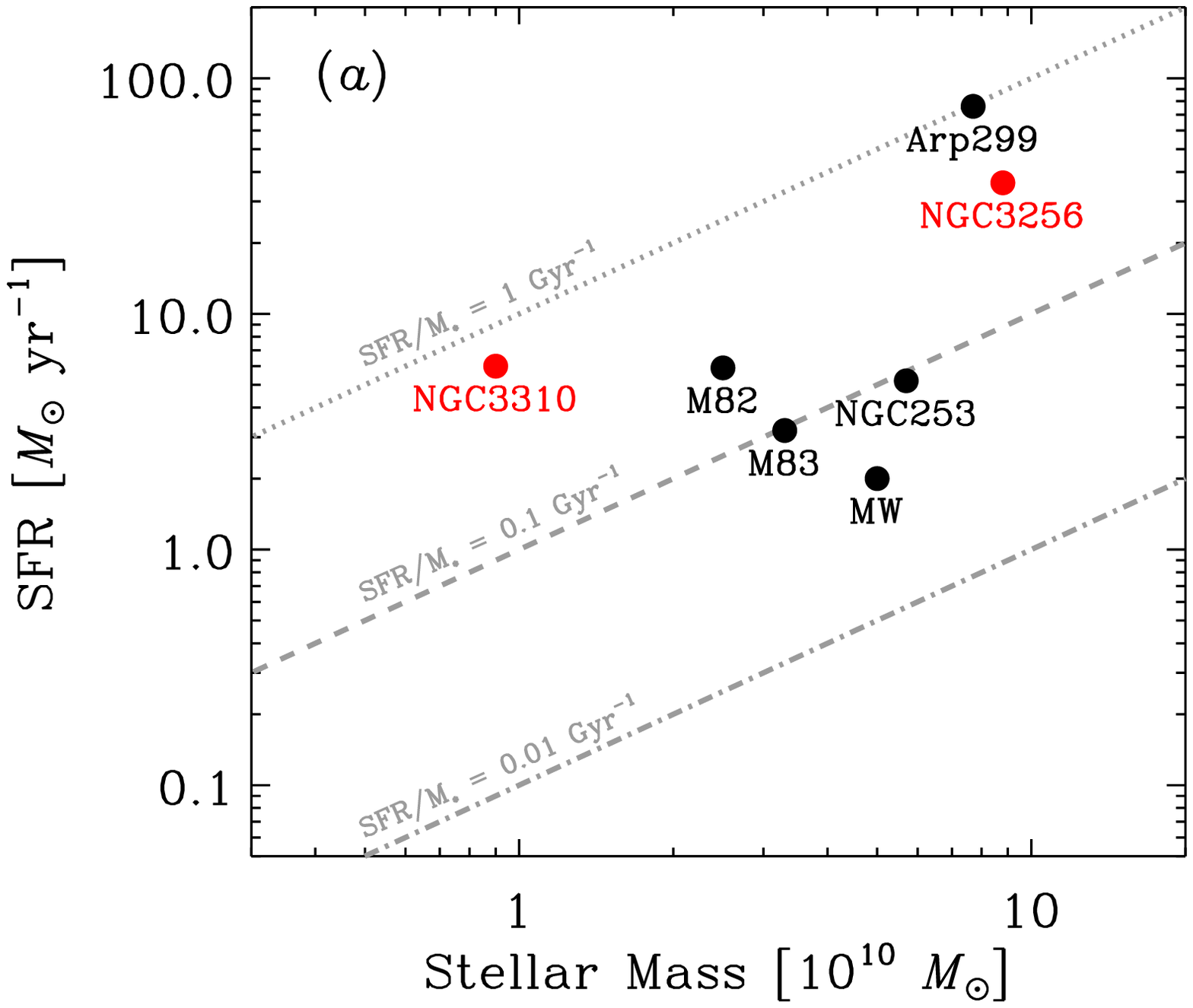}
\hfill
\includegraphics[width=9.2cm]{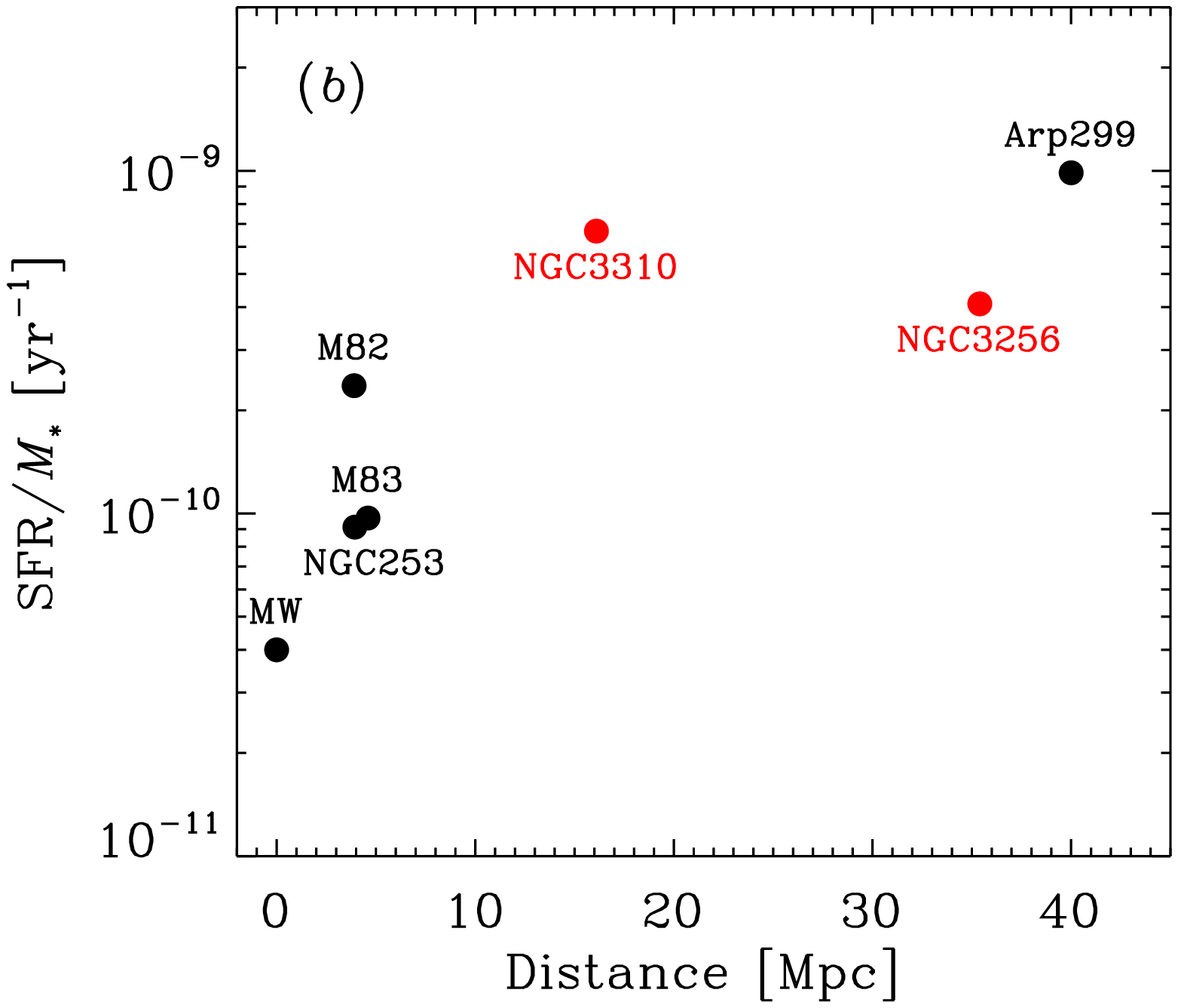}
}
\vspace{0.1in}
\caption{
Physical properties of \nustar\ starburst galaxies sample. ($a$) SFR
versus $M_\star$ for the full sample with regions of constant specific SFR
(sSFR~$\equiv$~SFR/$M_\star$) highlighted.  ($b$) sSFR versus distance for the
full sample.  For comparison, the location of the Milky Way (MW) has been
indicated.  Our sample consists of galaxies with sSFRs that range from being
comparable to those of the MW to a factor of $\approx$20 times higher.  We
therefore expect that the \xray\ emitting point-source populations in our
sample will be skewed towards including richer populations of HMXBs compared to
the MW (e.g., Lehmer \etal\ 2010; Mineo \etal\ 2012a).  NGC~3256 and NGC~3310
are amongst the highest sSFR galaxies in our sample.
}
\end{figure*}

With the launch of the \nustar\ \xray\ observatory (Harrison \etal\ 2013), we
now have a first-ever means for studying populations of normal galaxies (not
hosting luminous active galactic nuclei [AGN]) in the $\approx$10--30~keV
bandpass, an energy regime which is expected to be dominated by bright \xray\
binaries and low-level contributions from diffuse inverse-Compton emission
(e.g., Persic \& Rephaeli~2002; Wik \etal\ 2014a).  The spectra of \xray\
binaries at these energies betray unique information about their accretion
states as variations in accretion disk, reflection, and coronal components
undergo strong spectral pivots above $\approx$10~keV (see, e.g., Done \etal\
2007), a regime that was not previously probed by \chandra\ and \xmm\ for
sources outside the Local Group.  Prior to \nustar, studies of the hard \xray\
spectra of \xray\ binaries was limited to objects in the Milky Way (MW) and
Magellanic Clouds (e.g., via \rxte; see Remillard \& McClintock~2006 for a
review), providing limited information about how the accretion states of these
populations vary with galaxy properties (e.g., the starburst environment).  

With \chandra\ and \xmm, the lower-energy \hbox{$<$10~keV} emission from local
galaxies has been studied in detail for a variety of galaxy types (see, e.g.,
Fabbiano \etal\ 2006 for a review), and ultradeep \chandra\ observations have
extended these studies to very distant galaxy populations with look-back times
that span the vast majority of cosmic history (e.g., Basu-Zych \etal\ 2013a).
We are detecting \xray\ emission from galaxy populations out to $z \approx 4$
where the observed-frame \chandra\ 0.5--8~keV band corresponds to rest-frame
energies of 2.5--40~keV, the most sensitive regimes of the \nustar\ bandpass
for extragalactic binary work.  Results from such studies have prompted the
development of accreting binary population synthesis models, which have
successfully been employed to construct a self-consistent framework that
describes the observed evolution of \xray\ binary populations from cosmic dawn
to today (e.g., Fragos \etal\ 2013a,b).  Key to these observational
interpretations and model predictions, however, is knowledge of the \xray\
spectra at $\simgt$10~keV for galaxy populations that cover a broad range of
physical properties (e.g., Fragos \etal\ 2013a; Kaaret \etal\ 2014).
Therefore, a next step to improving our understanding of \xray\ binary
populations is to constrain the distributions and duty cycles of their
accretion states and measure how their resulting $\simgt$10~keV spectral
contributions vary with galaxy physical properties (e.g., star-formation rate
[SFR] and stellar mass [$M_\star$]).

We are conducting a joint \nustar\ and \chandra\ survey of six far-infrared
bright starburst galaxies (NGC~253, M82, M83, Arp~299, NGC~3256, and
NGC~3310) with the goal of quantifying the dominant processes that contribute to the
$\simgt$10~keV emission.  Figure~1 displays the SFR versus 
$M_\star$ and specific SFR (sSFR~$\equiv$~SFR/$M_\star$) versus distance
planes for the full galaxy sample.  Our key goals are to: (1) identify the
accretion states of the most luminous \xray\ binary populations in these
galaxies and characterize the $\approx$0.3--30~keV spectral energy
distributions (SEDs) as a function of the galaxy physical properties (e.g., SFR
and $M_\star$); (2) search for heavily obscured AGN that may be present in
these actively star-forming galaxies; and (3) constrain the nature of inverse
Compton emission associated with particle accelerations in starburst flows.

As a pilot program, we studied with \nustar\ and \chandra\ the nuclear region
and galaxy-wide \xray\ emission of the nearby starburst galaxy NGC~253 (Lehmer
\etal\ 2013; Wik \etal\ 2014a).  These studies showed that there was no evidence
for powerful buried AGN activity in the nucleus, and that the galaxy-wide
\xray\ emission above 10~keV is dominated by a few ultraluminous \xray\ sources
(ULXs) with minority contributions from lower luminosity \xray\ binaries that
have \nustar\ colors similar to Galactic black hole binaries in
intermediate accretion states.  The galaxy-wide \xray\ spectrum of NGC~253
steepens at energies above $\approx$6~keV, consistent with the spectra of
other ULXs studied by \nustar\ (see, e.g., Bachetti \etal\ 2013; Rana \etal\
2014; Walton \etal\ 2013, 2014), signaling a possible dominance of
super-Eddington accreting objects in the overall starburst galaxy spectra.

More recently, we have executed joint \nustar\ and \chandra/\xmm\ observations
of M83 (Yukita \etal\ in-preparation) and Arp~299 (Ptak \etal\ 2014), galaxies
with sSFRs that are comparable to and $\approx$20 times higher than that of the
MW, respectively.  The brightest \xray\ sources in M83 have \nustar\ colors
similar to those of NGC~253, albeit with fewer ULXs.  Arp~299 has
$\simgt$10~keV emission dominated by a single Compton-thick AGN that outshines
the other \xray\ emitting components of the galaxy.  Given the relatively large
distance ($\approx$40~Mpc) to Arp~299, and its correspondingly smaller angular
extent, it was not possible to spatially resolve the \xray\ emitting components
with \nustar.  Therefore, the spectral properties of the non-AGN components
were poorly constrained at energies above $\approx$10~keV, leaving us with
little knowledge of the high-energy spectral shape for high sSFR galaxies (see
Fig.~1b).

In this paper, we continue to investigate the 0.3--30~keV SEDs of 
starburst galaxies by studying two powerful starburst galaxies in the local
universe: NGC~3256 and NGC~3310. The sSFRs of these galaxies are the second and
third highest of our sample (see Fig.~1), and are
only exceeded by Arp~299.  Both NGC~3256 and NGC~3310 have been studied
extensively across the full electromagnetic spectrum, and neither system
exhibits compelling evidence for harboring powerful AGN (see $\S$2 for a
discussion). 
The sSFRs of these galaxies are $\approx$0.4--0.8~Gyr$^{-1}$, a range expected
to host \xray\ binary populations that are dominated by high-mass \xray\
binaries (HMXBs; e.g., Colbert \etal\ 2004; Lehmer \etal\ 2010). Such high-sSFR
galaxies are representative of the typical galaxy populations on the ``main
sequence'' at \hbox{$z \approx$~1--2} (e.g., Karim \etal\ 2011).  As such, this study
will provide SED constraints on high-sSFR galaxies that can be used to inform
\xray\ studies of high-redshift galaxy populations at these redshifts, which
require informed $K$-corrections (e.g., Lehmer \etal\ 2008; Basu-Zych \etal\
2013a; Kaaret~2014).

Throughout this paper, we assume distances of \hbox{$D = 35.4$~Mpc} and
19.8~Mpc (Sanders \etal\ 2003) and column densities of $N_{\rm H} = 9.6 \times 10^{20}$~cm$^{-2}$
and $1.1 \times 10^{20}$~cm$^{-2}$ for NGC~3256 and NGC~3310, respectively
(Dickey \& Lockman~1990).  Star-formation rates and stellar masses quoted
throughout this paper were calculated assuming a Kroupa~(2001) initial mass
function.  Star-formation rates were calculated using equation~6.11 of
Calzetti~(2013), making use of UV and mid-IR data from \galex\ and \spitzer,
respectively.  Stellar masses were estimated following the prescription
outlined in Appendix~2 of Bell \etal\ (2003), making use of $B-V$ colors from
RC3 (de~Vaucouleurs \etal\ 1991) and $K$-band luminosities based on 2MASS
photometry (Jarret \etal\ 2003) and the adopted distances.  These assumptions
imply SFR~$\approx$~36~\sfr\ and 6~\sfr\ and $M_\star \approx 9 \times
10^{10}$~\msol\ and $9 \times 10^{9}$~\msol\ for NGC~3256 and NGC~3310,
respectively.  Quoted errors associated with spectral fits represent 90\%
confidence intervals.  

%
\section{The Galaxies}
%

\subsection{NGC~3256}

NGC~3256 (35.4~Mpc) is a major-merger system composed of two gas-rich galaxies
with comparable masses (total $M_\star \approx 10^{11}$~\msol) that are in a
nearly coalescent phase of the merger (L{\'{\i}}pari \etal\ 2000;
Alonso-Herrero \etal\ 2002).  Two discrete nuclei have been identified
$\approx$5~arcsec apart (850~pc; Zenner \& Lenzen~1993; Norris \& Forbes~1995)
and remarkably long (200~kpc) H~{\small I} tidal features have been identified
(e.g., Graham \etal\ 1984; English \etal\ 2003).  The system has the highest
\hbox{8--1000~$\mu$m} infrared luminosity ($L_{\rm IR} \approx$~$4 \times 10^{11}$~$L_{\odot}$;
Sanders \etal\ 2003) for galaxies that reside within $z \simlt 0.01$, and
studies at radio (e.g., Norris \& Forbes~1995), submm (e.g., Sakamoto \etal\
2006, 2014), infrared (Graham \etal\ 1984; Doyon \etal\ 1994a, 1994b; Moorwood
\& Olivia~1994; Lira \etal\ 2008; Alonso-Herrero \etal\ 2013), optical (e.g.,
Heckman \etal\ 2000; L{\'{\i}}pari \etal\ 2000; Alonso-Herrero \etal\ 2002),
and UV (e.g., Kinney \etal\ 1993; Leitherer \etal\ 2013) wavelengths all
support a scenario in which the majority of the galaxy's power output is
produced by starburst-related activity.  

X-ray studies with \chandra\ and \xmm\ have shown that the 0.3--10~keV spectrum
of NGC~3256 can be modeled well as a multiphase thermal plasma with harder
power-law emission from a population of discrete point sources, the majority of
which are ULXs (see, e.g., Lira \etal\ 2002;
Jenkins \etal\ 2004; Pereira-Santaella \etal\ 2011).  Although both nuclear
regions have been detected by \chandra\ with $L_{\rm 2-10~keV} \sim
10^{40}$~\lum, the nature of the \xray\ emission is unclear, as no obvious AGN
signatures have been found (Lira \etal\ 2002).  Detailed assessment of the
kinematics of the molecular gas, via ALMA and SMA observations, have concluded
that both nuclei contain powerful outflows, but with differing characteristics
(e.g., Cicone \etal\ 2014; Sakamoto \etal\ 2014; Emonts \etal\ 2014).  The
northern nuclear outflow is of wide breadth and has a mean velocity of a
few~100~km~s$^{-1}$ and is likely driven by a starburst superwind.  The
southern outflow is well-collimated and faster ($\simgt$1000~km~s$^{-1}$) and
is plausibly driven by an AGN that has recently gone dormant (e.g., Moran
\etal\ 1999; Neff \etal\ 2003; Alonso-Herrero \etal\ 2012).

\subsection{NGC~3310}

%
%
\begin{figure}
\figurenum{2}
\centerline{
\includegraphics[width=8.9cm]{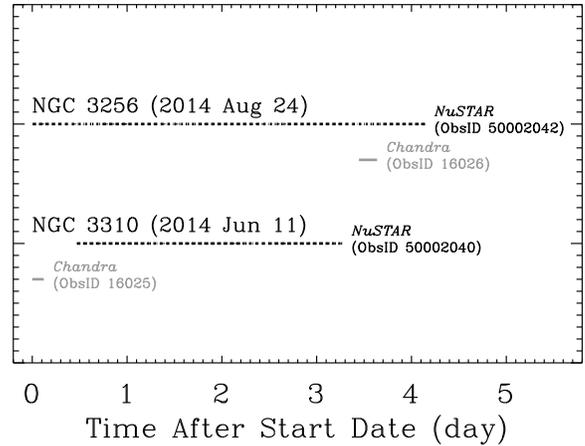}
}
\vspace{0.1in}
\caption{
Relative \nustar\ ({\it black lines\/}) and \chandra\ ({\it gray lines})
observational coverage for NGC~3256 ({\it top\/}) and NGC~3310 ({\it
bottom\/}).  For clarity, we have annotated the starting date of the observational
epoch.  The apparently broken up \nustar\ observational intervals are due
primarily to Earth occultations and passages through the SAA, which result in
observing efficiencies of $\approx$51--58\% on average.
}
\end{figure}

%
%
\begin{figure*}
\figurenum{3}
\centerline{
\includegraphics[width=8.5cm]{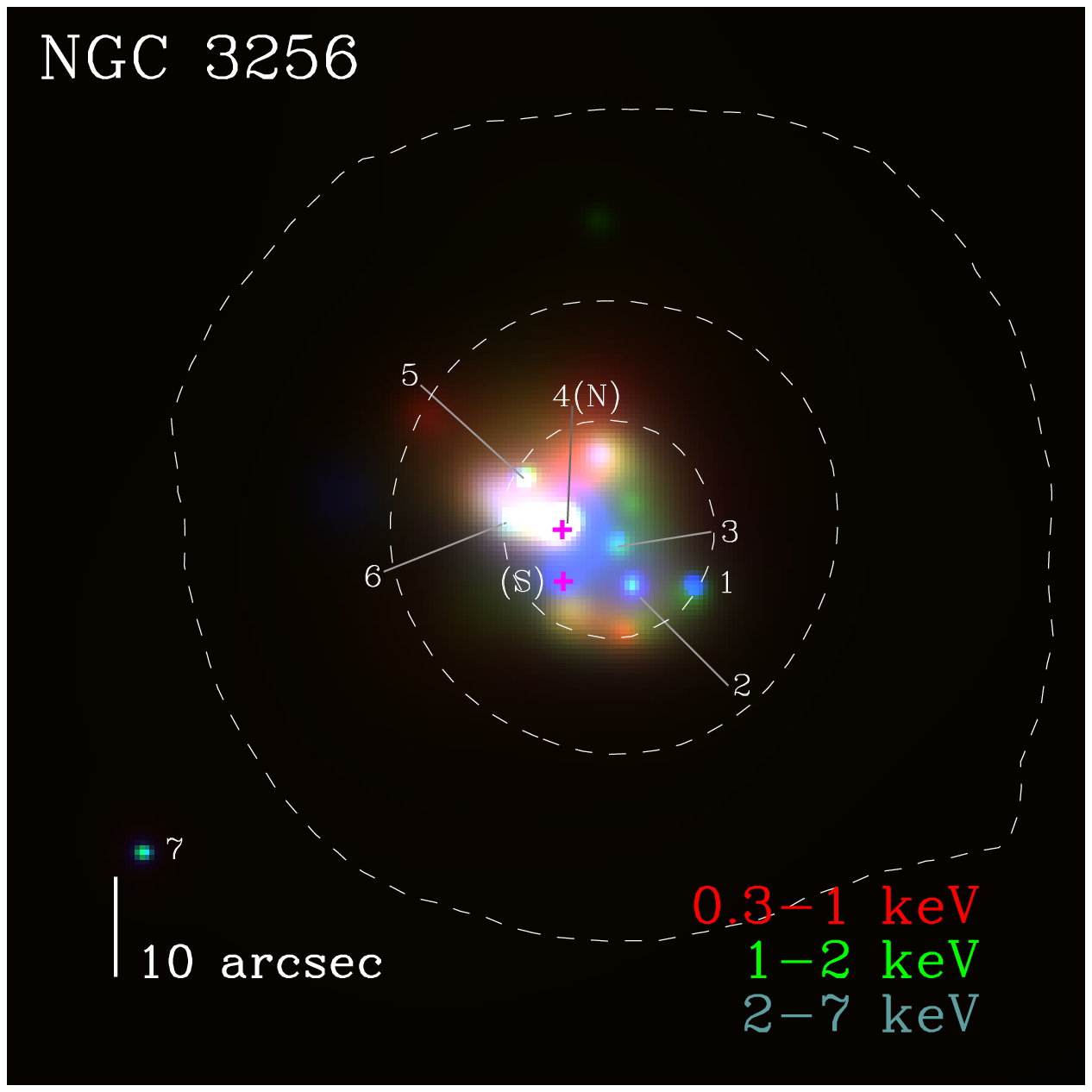}
\hfill
\includegraphics[width=8.5cm]{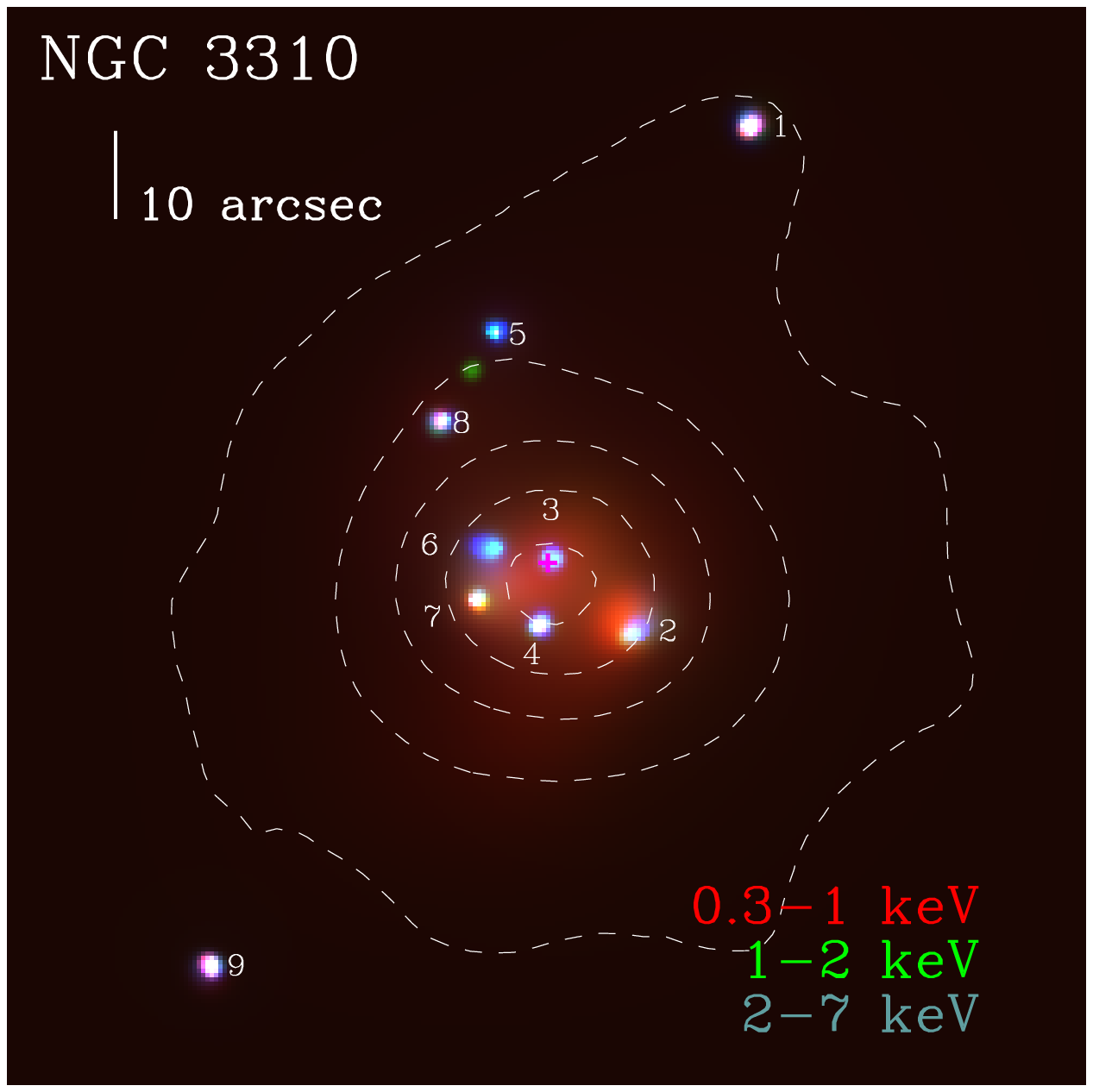}
}
\vspace{0.1in}
\caption{
Three-color adaptively smoothed \chandra\ images of NGC~3256 and NGC~3310 with
4--25~keV \nustar\ intensity contours overlaid ({\it dashed curves\/}).  The three bandpasses,
0.3--1~keV ($red$), 1--2~keV ($green$), and 2--7~keV ($blue$) were chosen to
highlight the diffuse components from hot gas and unresolved point sources, as well as
absorption in the bright point-source populations.  A 10~arcsec bar has been
added to each figure for reference.  The annotated source numbers, ordered by increasing right ascention, highlight
bright point sources that are studied in $\S$4.3 and presented in Table~2.  The
locations of galactic nuclei are indicated with magenta crosses.  For NGC~3256,
the northern (N) and southern (S) nuclei are labeled.  With \chandra, we detect
the northern nucleus of NGC~3256 (source~4, {\it left\/}) and the nuclear
region of NGC~3310 (source~3, {\it right\/}).   The \nustar\ FWHM is
$\approx$18~arcsec, which is sufficient for resolving some structure in the
nearer NGC~3310; however, the \nustar\ emission for NGC~3256 is consistent with
a single source PSF.
}
\end{figure*}

NGC~3310 is a closer (19.8~Mpc) starburst system with a SFR that is a
factor of $\sim$10 times lower than NGC~3256 ($L_{\rm IR} \approx 4
\times 10^{10}$~$L_{\odot}$; Sanders \etal\ 2003); however, given its
relatively low stellar mass ($M_\star \simlt 10^{10}$~\msol), this system has a
comparable and perhaps larger sSFR than NGC~3256 (see Fig.~1).  The system
contains evidence for a minor-merger that has triggered a vigorous, young
($\approx$2.5--5~Myr old) circumnuclear star-forming ring with a
$\approx$20--30~arcsec ($\approx$2~kpc) diameter (e.g., Balick \& Heckman~1981;
D{\'{\i}}az \etal\ 2000; Elmegreen \etal\ 2002; Chandar \etal\ 2005;
Miralles-Caballero \etal\ 2014).  The most recent minor merger was inferred to
occur in the last $\approx$30~Myr (Elmegreen \etal\ 2002; de~Grijs \etal\
2003a), and is likely to be one of several past interactions with small
metal-poor dwarf galaxies (e.g., Wehner \etal\ 2006), as evidence for starburst
activity as old as 100~Myr is apparent in the star clusters (Meurer~2000).
Such a series of minor-mergers could potentially provide a significant increase
to the mass build-up of the bulge (Miralles-Caballero \etal\ 2014) and is
likely to have modified the gas-phase abundances, as there is evidence that the
circumnuclear metallicity is lower than is typical for spiral galaxies with
morphologies similar to NGC~3310 (e.g., Pastoriza \etal\ 1993).

Similar to NGC~3256, the \chandra\ and \xmm\ spectra of NGC~3310 can be modeled
well by a multiphase hot gas plus a power-law component; however, the plasma
temperature is cooler and the power-law slope is harder than those found for
NGC~3256 (Jenkins \etal\ 2004).  The hard power-law component is dominated by a
population of $\approx$9 ULXs that are distributed within the circumnuclear
star-forming ring and along the spiral arms. \hst\ STIS observations of the
nuclear region have uncovered a bright $\approx$0.5~arcsec ($\approx$40~pc)
diameter central star-forming region with kinematic properties consistent with
a circularly rotating disk (Pastorini \etal\ 2007).  The nuclear region is
detected by \chandra\ as a point-like source,\footnote{The nuclear region is
also detected by \xmm; however, it is clear that confusion with four other
\chandra-resolved ULXs of comparable brightness makes it difficult to uniquely interpret those data.}
and the luminosity of this region, $L_{\rm 2-10~keV} \sim 10^{40}$~\lum, is
within the range of other ULXs in the galaxy.  The \chandra\ spectrum shows
that the nuclear \xray\ source has a flat power-law spectral slope, with some evidence
(at the 2$\sigma$ confidence level) for an Fe~K$\alpha$ line---two properties
that indicate there may be a hidden (obscured) AGN in the galaxy nucleus
(Tzanavaris \etal\ 2007).

%
\section{Observations and Data Reduction}
%

In this investigation we make use of nearly simultaneous observations with
\nustar\ and \chandra\ to explore the full 0.3--30~keV bandpass properties.  In
later sections, we make use of archival \chandra\ data to support our analyses;
however, we focus the current investigation on the new nearly simultaneous
observations.  We obtained our \nustar\ and \chandra\ observations of NGC~3256
and NGC~3310 over single epochs beginning on 2014~Aug~24 and 2014~Jun~11,
respectively.  Figure~2 shows the relative observing schedules for the \nustar\
and \chandra\ exposures.  Due to Earth occultations and passages through the
South Atlantic Anomaly (SAA), the \nustar\ on-target observations were carried
out at \hbox{$\approx$51--58\%} efficiency over \hbox{$\approx$3--4~day}
periods.  The cumulative \nustar\ good-time-interval exposures were 184~ks
(ObsID 50002042) and 141~ks (ObsID 50002040) for NGC~3256 and NGC~3310,
respectively.  The \chandra\ observations were continuous and lasted 16~ks
(ObsID 16026) and 10~ks (ObsID 16025) for NGC~3256 and NGC~3310, respectively.  

%
%
\begin{figure*}
\figurenum{4}
\centerline{
\includegraphics[width=8.6cm]{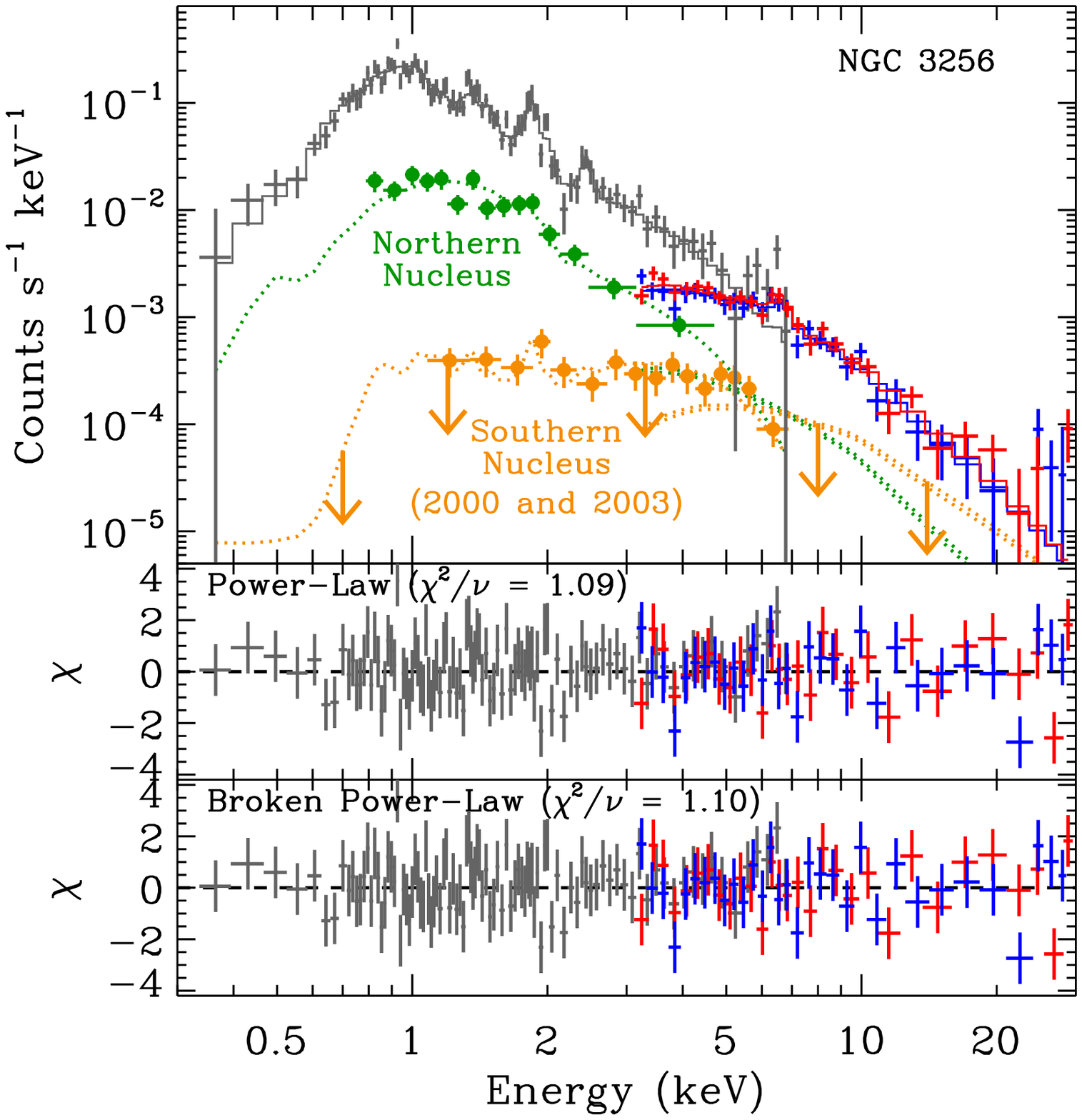}
\hfill
\includegraphics[width=8.6cm]{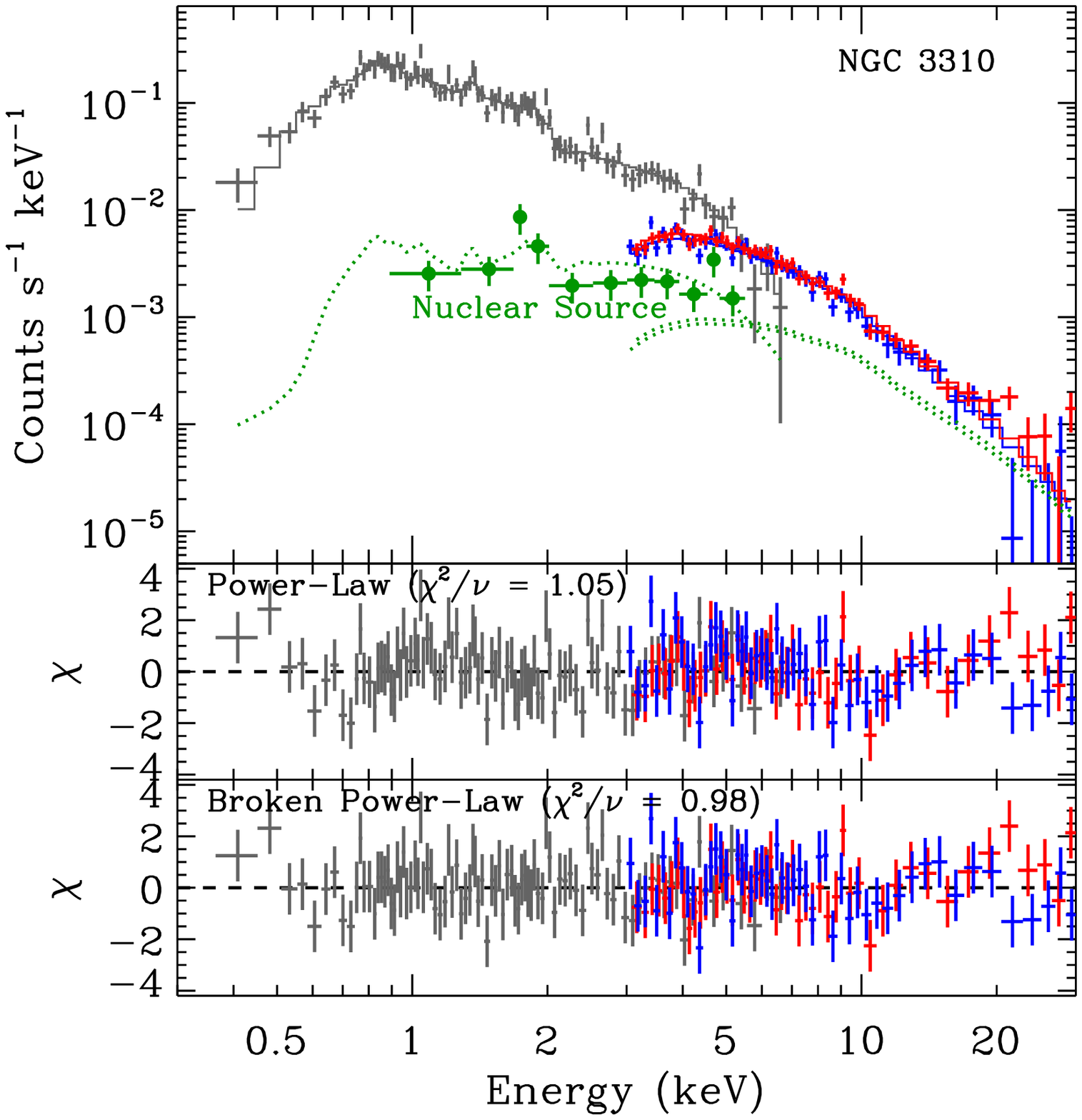}
}
\vspace{0.1in}
\caption{
Joint nearly simultaneous \chandra\ ({\it gray\/}) and \nustar\ ({\it red} for
FPMA and {\it blue} for FPMB) spectra for the full galactic extents of NGC~3256
({\it left\/}) and NGC~3310 ({\it right\/}).  The diffuse emission that is
spatially isolated with \chandra\ was used to constrain two-temperature thermal
({\ttfamily apec}) models for both galaxies.  For NGC~3256, an additional
obscured thermal component ($\approx$1~keV) was added to account for hot gas
emission associated with the nuclear starburst.  Absorbed single and broken
power-law models that account for the \chandra-detected binaries were both
tested for each galaxy, and the resulting residuals are displayed, and
goodness-of-fit values are annotated, in the bottom two panels of each plot.
We find that a single power-law is sufficient to fit the non-thermal emission
from NGC~3256, while a broken power-law is preferred in the case of NGC~3310.
We also display the contributions from nuclear components of these galaxies,
with extrapolations to the \nustar\ band.  For NGC~3256, this includes
contributions from two nuclear regions associated with two galaxies in the
merger: the northern nucleus contribution ({\it green curve\/}) and the
estimated upper limit to the southern nucleus contribution ({\it orange
curve\/}).  The southern nucleus was not formally detected in our 2014
observation, and the upper limit spectrum here indicates the average \chandra\
spectrum from deeper observations in 2000 and 2003 scaled to the upper limit
flux of our 2014 observation. We note that the southern nuclear region of
NGC~3256 and the nuclear region of NGC~3310 both have flat \xray\ spectra and
are candidate low luminosity AGN (see $\S$4.3 for details).
}
\end{figure*}

The \nustar\ data were reduced using {\ttfamily HEASoft} v6.15, \nustar\ Data
Analysis Software ({\ttfamily NuSTARDAS}) v1.7, and CALDB version 20130320.  We
processed level 1 data to level 2 products by running {\ttfamily nupipeline},
which performs a variety of data reduction steps, including (1) filtering out
bad pixels, (2) screening for cosmic rays and observational intervals when the
background was too high (e.g., during passes through the SAA), and (3)
projecting accurately the events to sky coordinates by determining the optical
axis position and correcting for the dynamic relative offset of the optics
bench to the focal plane bench due to motions of the 10~m mast that connects
the two benches.

Due to the angular extents of NGC~3256 and NGC~3310 (major-axes $2a$
=~3.8~arcmin and 3.1~arcmin, respectively), the \chandra\ exposures for both
galaxies were conducted using single ACIS-S pointings with the approximate
position of the galactic centers set as the aimpoints.  For our data reduction,
we used {\ttfamily CIAO}~v.~4.6 with {\ttfamily CALDB}~v.~4.6.1.1.  We
reprocessed our events lists, bringing level~1 to level~2 using the script
{\ttfamily chandra\_repro}, which identifies and removes events from bad pixels
and columns, and filters events lists to include only good time intervals
without significant flares and non-cosmic ray events corresponding to the
standard \asca\ grade set (grades 0, 2, 3, 4, 6).  We constructed an initial
\chandra\ source catalog by searching the \hbox{0.5--7~keV} images with
{\ttfamily wavdetect} (run with a point spread function [PSF] map created using
{\ttfamily mkpsfmap}), which was set at a false-positive probability threshold
of $2 \times 10^{-5}$ and run over seven scales from 1--8 (spaced out by
factors of $\sqrt{2}$ in wavelet scale: 1, $\sqrt{2}$, 2, 2$\sqrt{2}$, 4,
4$\sqrt{2}$, and 8).  Using these initial \chandra\ source catalogs, we
constructed source-free light curves and searched for flaring intervals that
were 5$\sigma$ above the background level.  No such intervals were found, and
our observations were deemed to be sufficiently cleaned.

In Figure~3, we show the adaptively smoothed, false color \chandra\ images of
NGC~3256 and NGC~3310, with \nustar\ \hbox{4--25~keV} intensity contours
overlaid.  It is apparent from the \chandra\ data that diffuse emission at $E <
1$~keV, potentially due to hot gas and unresolved point sources, and multiple
bright point sources at all energies provide dominant contributions to the
\chandra-detected emission for both galaxies.  The \nustar\ PSF has an
18~arcsec full width half maximum (FWHM) core with a 58~arcsec half-power
diameter (Harrison \etal\ 2013).  Given the \xray\ extents of the galaxies
(60--70~arcsec), \nustar\ is unable to strictly resolve the point-source
populations that dominate the $E > 3$~keV emission probed by \nustar; however,
some evidence for spatial extent following the point-source distribution is
apparent in NGC~3310 (see \S4.3 below).

%
\section{Analysis and Results}
%

As discussed in $\S$1, our primary goals are to measure and assess the contributing
components to the broad-band \hbox{0.3--30~keV} SEDs of the two powerful
starburst galaxies NGC~3256 and NGC~3310, and constrain the underlying AGN
activity in these galaxies.  To achieve these goals, we first characterized the
\hbox{0.3--30~keV} spectra across the full extents of both galaxies.  We then
used the high spatial resolution of \chandra\ to measure the $\simlt$8~keV
contributions from bright resolved point-sources and diffuse emission from hot
gas and unresolved point-sources.  Finally, using sensible extrapolations of
the \chandra\ component spectra to $\simgt$8~keV, we infer the relative
contributions AGN candidates would make to the \nustar\ galaxy-wide
spectra.  In the sections below, we describe each of these procedures in
detail.


\begin{table*}
\begin{center}
\caption{Best fit parameters for full-galaxy spectra of NGC~3256 and NGC~3310.}
\begin{tabular}{lccc}
\hline\hline
\multicolumn{1}{c}{\sc Parameter} & {\sc Unit} & {\sc NGC~3256} & {\sc NGC 3310} \\
\hline
%
\multicolumn{4}{c}{\chandra\ only fit:$^\ast$ point-source subtracted diffuse emission} \\
\hline
$N_{\rm H, Gal}$ \dotfill          &   10$^{22}$ cm$^{-2}$                              & 0.096                   & 0.011 \\
$kT_1$ \dotfill                    &   keV                                              & $0.33^{+0.06}_{-0.04}$  & $0.2 \pm 0.1$  \\
$N_{\rm H,2}$ \dotfill                 &   10$^{22}$ cm$^{-2}$                              & $0.74^{+0.09}_{-0.04}$  & $0.6^{+0.2}_{-0.1}$  \\
$kT_2$ \dotfill                    &   keV                                              & $0.90^{+0.06}_{-0.08}$  & $0.28^{+0.08}_{-0.08}$  \\
\hline
\multicolumn{4}{c}{{\it Chandra} + {\it NuSTAR} galaxy-wide fit:$^\dagger$ diffuse emission plus non-thermal point sources} \\
\hline
$N_{\rm H,3}$ \dotfill             &   10$^{22}$ cm$^{-2}$                              & $115^{+178}_{-93}$             & \ldots \\
$kT_3$ \dotfill                    &   keV                                              & $1.0^{+2.4}_{-0.4}$           & \ldots  \\
$N_{\rm H, PL}$ \dotfill           &   10$^{22}$ cm$^{-2}$                              & $3.0^{+1.9}_{-1.2}$           & $0.89^{+0.30}_{-0.56}$ \\
$\Gamma_1$    \dotfill             &                                                    & \ldots                        & $2.07^{+0.23}_{-0.59}$ \\
$E_{\rm break}$   \dotfill         &   keV                                              & \ldots                        & $4.70^{+0.72}_{-1.21}$ \\
$\Gamma_2$    \dotfill             &  \phantom{photons keV$^{-1}$~cm$^{-2}$~s$^{-1}$ at 1 keV blah blah}  & $2.64 \pm 0.20$   & $2.65^{+0.10}_{-0.14}$ \\

$A_{kT_1}$  \dotfill      &      & $(1.77 \pm 0.20) \times 10^{-4}$           & $(1.15 \pm 0.32) \times 10^{-4}$ \\
$A_{kT_2}$  \dotfill      &      & $(1.29^{+0.08}_{-0.10}) \times 10^{-3}$           & $2.85^{+0.28}_{-0.55} \times 10^{-4}$ \\
$A_{kT_3}$  \dotfill      &      & $4.0^{+1046}_{-4.0} \times 10^{-2}$     & \ldots \\
$A_{\rm PL}$  \dotfill        &   photons keV$^{-1}$~cm$^{-2}$~s$^{-1}$ at 1 keV   & $4.6^{+1.8}_{-2.5} \times 10^{-4}$     & $7.4^{+2.9}_{-3.8} \times 10^{-4}$ \\

$\chi_{\rm red}^2$     \dotfill    &                                                    & 1.09                    & 0.98 \\
$\nu$        \dotfill              &   (degrees of freedom)                             & 159                     & 192 \\
\hline
\hline
$f_{\rm X}$ (10--30~keV) \dotfill  &   10$^{-13}$ ergs cm$^{-2}$ s$^{-1}$               & 1.33$\pm$0.03               & 5.9$\pm$0.4 \\
$L_{\rm X}$ (10--30~keV)  \ldots   &   10$^{40}$ ergs s$^{-1}$                          & 1.99$\pm$0.05               & 2.8$\pm$0.2 \\
\hline
\end{tabular}
\end{center}
{\bf Note---All quoted errors are at the 90\% confidence level.}\\
$^\ast$Fits to the point-source subtracted diffuse \chandra\ spectra were performed using {\ttfamily XSPEC} model {\ttfamily tbabs$_{\rm Gal}*$(apec$_1$+tbabs$_2*$apec$_2$+tbabs$_{\rm PL}*$pow}) to obtain values of $T_1$, $T_2$, and $N_{\rm H,2}$.  In this process the power-law had a fixed photon index of $\Gamma = 1.8$ and all thermal models assumed solar abundances.  Values of $T_1$, $T_2$, and $N_{\rm H,2}$ were subsequently fixed when fitting the \chandra\ + \nustar\ galaxy-wide spectra, where their normalizations were free to vary.\\
$^\dagger$ The galaxy-wide \chandra\ + \nustar\ spectra were fit using {\ttfamily XSPEC} models {\ttfamily tbabs$_{\rm Gal}*$(apec$_1$+tbabs$_2*$apec$_2$+tbabs$_3*$apec$_3$+tbabs$_{\rm PL}*$bknpo}) and {\ttfamily tbabs$_{\rm Gal}*$(apec$_1$+tbabs$_2*$apec$_2$+tbabs$_{\rm PL}*$pow}) for NGC~3256 and NGC~3310, respectively. \\
\end{table*}

\subsection{Galaxy-Wide Spectral Analyses}

We began by extracting the nearly simultaneous \chandra\ and \nustar\ spectra
over the full extents of both galaxies.  For NGC~3256 and NGC~3310, we
extracted on-source spectra using circular apertures with radii $r_{\rm src}$
equal to 80~arcsec and 90~arcsec, respectively.  These apertures were chosen to
encompass the entire optical extents of the galaxies and reach 20~arcsec
(roughly the \nustar\ FWHM) beyond the most offset point-sources that were
clearly detected with \chandra.  For each galaxy, background spectra were
extracted using \hbox{1--5} circular apertures located in source-free regions.
For \chandra, the background extraction regions had radii
$\approx$50--70~arcsec and were chosen by eye to be placed in a pattern
surrounding the on-source extraction region.  For \nustar, background regions
were chosen more carefully to properly account for the spatial background
gradients that arise primarily from the ``aperture'' background component,
which contains cosmic \xray\ background stray light that shines directly onto
the detectors from a $\approx$1--4~deg annular region (see, e.g., Appendix~A of
Wik \etal\ 2014b).  Following the procedure described in Appendix~B of Wik
\etal\ (2014b), we constructed background maps for the FPMA and FPMB modules in
the \hbox{3--20~keV} band, an energy regime sensitive to variations in the
aperture background component.  For each background map, we identified 1--4
regions with sizes equal to the on-source aperture that had background levels
comparable to those expected to be present within the source extraction
regions.  The closest edge of each background region was located $>$$1.2\times
r_{\rm src}$ away from the central positions of each galaxy to avoid
contamination from the galaxy itself.  After choosing appropriate regions, the
\chandra\ and \nustar\ on-source and background spectra were extracted using
the {\ttfamily specextract} and {\ttfamily nuproducts} tools, respectively.

In Figure~4, we show the 0.3--30~keV galaxy-wide spectra for both galaxies.  As
discussed in \S3, from the \chandra\ data (see Fig.~3), it is clear that the
\xray\ emission from both galaxies can be broadly characterized as consisting
of diffuse emission (from hot gas and unresolved point sources) and bright
point sources (e.g., \xray\ binaries and possibly AGN).  Previous
investigations with \chandra\ and \xmm\ have found that the diffuse components of
these galaxies can be modeled well using two-temperature thermal spectra, with
NGC~3310 having cooler temperatures ($kT$~=~0.3 and 0.6~keV) than the more
powerful NGC~3256 ($kT$~=~0.6 and 0.9~keV), which may also contain an
additional absorbed hot ($\approx$3.9~keV) component (e.g., Lira \etal\ 2002;
Jenkins \etal\ 2004).  The brightest point sources in these galaxies are ULXs
with nine sources spanning $L_{\rm X} \approx$~\hbox{(2--10)}~$\times 10^{39}$~\lum\
for NGC~3310 (Smith \etal\ 2012) and roughly a dozen sources with $L_{\rm X}
\approx$~(2--60)~$\times 10^{39}$~\lum\ in NGC~3256 (Lira \etal\ 2002).   These
ULXs dominate the galaxy-wide emission from \hbox{$\approx$2--8~keV} and are
expected to provide majority contributions to the non-AGN emission in the
$\approx$8--30~keV bandpass.   We therefore chose to model the broad-band
\hbox{0.3--30~keV} spectra of the galaxies using the sum of thermal and
power-law components to account for the hot gas and ULXs, respectively.

For each galaxy, we tested for the presence of \hbox{2--3} thermal components,
in which the absorption of each component increases with increasing
temperature.  Such a trend has been seen in spatially resolved starburst
galaxies like M82, NGC~253, and past studies of NGC~3256, in which the hottest,
most obscured components are located in the nuclear regions and the cooler,
unobscured plasmas extend to larger galaxy-wide scales (e.g., Strickland \etal\
2000; Pietsch \etal\ 2001; Lira \etal\ 2002; Strickland \& Heckman 2007, 2009;
Ranalli \etal\ 2008).  We made use of the spatial resolving power of \chandra\
to better inform our estimates of the diffuse emission component.  To this end,
we extracted \chandra\ ``diffuse component'' spectra from the galaxies after
removing the bright point-source populations.  In this exercise, we excluded
events from regions that were within 1.5~$\times$ the radius of the 90\%
encircled-energy fraction PSFs of all bright \chandra\ detected point sources
(see $\S$~4.2 below for details on the \chandra\ detected point-source
population) and extracted events from within the same apertures used for the
galaxy-wide spectra discussed above.  The resulting diffuse component spectra
are expected to be dominated by hot gas and unresolved, low luminosity \xray\
point-sources.  We note that the bright \xray\ point-source densities are
highest in the central regions of the galaxies where star-formation is highest.
It is in these same regions where we expect to have the hottest \xray-emitting
gas associated with the starbursts.  For NGC~3256 and NGC~3310, the removal of
these regions will therefore likely result in the removal of a significant
fraction of the emission from the hottest gas component.  

As a first step, we fit the diffuse, point-source-subtracted \chandra\ spectra
using two thermal components ({\ttfamily apec} in {\ttfamily XSPEC}; Smith
\etal\ 2001) plus a single power-law (accounting for unresolved point sources),
to determine the temperatures of the galaxy-wide hot gas.\footnote{We modeled
the point-source-subtracted \chandra\ spectra using {\ttfamily XSPEC} model
{\ttfamily tbabs*(apec + tbabs*apec + tbabs*powerlaw)} and fixed the initial
absorption to Galactic and the power-law spectral slope to $\Gamma = 1.8$, a
value consistent with that of bright \xray\ binaries (e.g., Mineo \etal\
2012a).  The {\ttfamily apec} component abundances were fixed to solar.} In the
top portion of Table~1, we list the best-fit temperatures that result from this
process.  Similar to the results presented by Jenkins \etal\ (2004), we find
best-fitting temperatures of $kT \approx$~0.3 and 0.9~keV for NGC~3256; 
however, we find that much cooler temperatures of $kT \approx$~0.2 and 0.3~keV
were required for NGC~3310, with the latter component being more absorbed and
intrinsically powerful than the former component.

Next, for each galaxy, we modeled the full galaxy-wide \chandra\ plus \nustar\
spectrum using the two thermal components (with temperatures fixed at the
previous best-fit values) plus a third, hotter ($kT \simgt$~1~keV) and more
obscured thermal component and a non-thermal component to model emission from
the point source population.  For the point source spectral component, we
tested both a single and broken power-law model to see if breaks (e.g., at $E
\approx$~3--8~keV) were preferred, as is most often the case for
individual ULX spectra (e.g., Gladstone \etal\ 2009; Bachetti \etal\ 2013;
Lehmer \etal\ 2013; Rana \etal\ 2014; Wik \etal\ 2014a).  Since the
point-source spectra of our galaxies contain the conglomerated summed emission
from primarily ULXs, we do not report fits using more complex models that
contain detailed physical treatment of individual accretion disks and coronae
(e.g., {\ttfamily diskbb}, {\ttfamily diskpbb}, {\ttfamily diskpn}, {\ttfamily
comptt}, {\ttfamily eqpair}, and {\ttfamily dkbbfth}).  In Table~1, we provide
the best-fitting parameters for fits to the \chandra\ plus \nustar\ spectra
that include the two temperature plasma (with fixed temperatures) component,
plus an absorbed hotter component (in the case of NGC~3256 only), plus an
absorbed single or broken power-law component.  In the bottom panels of
Figure~4 we display the residuals to these fits.

For NGC~3256, we find that the sum of the diffuse component plus a single
absorbed ($N_{\rm H} \approx 3 \times 10^{22}$~cm$^{-2}$) power-law ($\Gamma
\approx 2.6$) and an absorbed hot ($\approx$1~keV) plasma component were
sufficient to obtain a good fit to the \hbox{0.3--30~keV} spectrum.  The
absorbed hot plasma component was implemented to fit a notable Fe-line feature
in the \nustar\ spectrum at $\approx$6.4--6.8~keV.  Such a feature is also
notable in the \chandra\ spectrum of the full galaxy but not the diffuse region
with point sources removed (see above).  Further analysis indicates that the Fe
emission photons are concentrated in a region within $\approx$20~arcsec of the
nucleus, where most of the \xray\ detected point sources are located.  This
component was not in our diffuse, point-source-subtracted \chandra\ spectrum
because of the spatial coincidence with point sources. We note that \xray\
binaries and obscured AGN may also provide some contribution to such a feature.
However, given the steepness of the galaxy-wide spectral slope above
$\simgt$8~keV, it is unlikely that the Fe-line is powered by an obscured AGN,
which would have a harder spectral slope ($\Gamma \approx$~1.5--2.0; see
$\S$4.3).  Replacing the single power-law with a broken power-law does not
improve the quality or change the character of the fit (see lower panels of
Fig.~4 left).  Since the thermal components in the galaxy dominates the
$\approx$0.3--2.5~keV emission in NGC~3256, a true break in the power-law
component could be masked if it resides at $E \simlt 3$~keV.  

For NGC~3310, we find that the diffuse component plus a broken power-law model
provides the best fit to the data; no additional absorbed hotter
component was required to fit the spectrum, as we do not see clear evidence for
an Fe-line.  The best-fitting slopes ($\Gamma_1 \approx 2.1$ and $\Gamma_2
\approx 2.7$) and energy break ($E_{\rm break} \approx 4.7$~keV) are within the
range of values found for ULXs (e.g., Gladstone \etal\ 2009).  In contrast to
NGC~3256, which had strong emission features related to hot gas out to $E
\approx 3$~keV, we find that the overall spectrum of NGC~3310 above
$\approx$1~keV is dominated by the point-source population.  As we will discuss
in more detail below, the SFR-normalized point-source emission in NGC~3310 is a
factor of \hbox{$\approx$3--10} times higher compared to other starburst galaxies
studied in our program (i.e., NGC~253, NGC~3256, and M83).  The relatively strong emission
from the point-source population may therefore be masking lower-intensity line
features that would betray the presence of a hot \xray\ plasma or an obscured
AGN (e.g., Fe emission-lines).  If NGC~3310 harbors an obscured AGN, it is
almost certainly of low luminosity and is likely to provide only a minor
perturbation on the non-AGN emission in the \nustar\ band.  We return to the
discussion of the potential for AGN in both NGC~3256 and NGC~3310 in \S4.3
below when we discuss the \chandra\ properties of the nuclear sources in
detail.

\begin{table*}
\begin{center}
\caption{Chandra Point-Source Properties}
\begin{tabular}{lccccccccccccc}
\hline\hline
 & & & \multicolumn{3}{c}{{\ttfamily powerlaw}} & \multicolumn{3}{c}{{\ttfamily diskbb}} & & &  & & \\
 & & & \multicolumn{3}{c}{{\rule{1.0in}{0.01in}}} & \multicolumn{3}{c}{{\rule{1.0in}{0.01in}}} &  & &  & & \\
   & $\alpha_{\rm J2000}$ & $\delta_{\rm J2000}$ & $N_{\rm H, int}$ &   &   &   $N_{\rm H, int}$ & $kT_{\rm in}$  & &  &    $\log f_{\rm 2-10~keV}$ & $\log L_{\rm 2-10~keV}$  & Net Counts & \\
 \multicolumn{1}{c}{Source ID}  & (hr) & (deg) & (10$^{22}$~cm$^{-2}$) & $\Gamma$ & $C$ & (10$^{22}$~cm$^{-2}$) & (keV) & $C$ & $\nu$  & ($\log$ \flux) & ($\log$ \lum) & (0.5--7~keV) & Note \\
 \multicolumn{1}{c}{(1)}  & (2) & (3) & (4) & (5) & (6) & (7) & (8) & (9) & (10) & (11) & (12) & (13) & (14) \\
\hline\hline
 \multicolumn{14}{c}{NGC~3256}   \\
\hline
 1\dotfill &      10 27 50.0 &   $-$43 54 19.8 &    0.02$^{+0.38}_{-0.02}$ &       1.4$^{+0.9}_{-0.5}$ &   43 &                   $<0.16$ &       1.9$^{+4.7}_{-0.7}$ &  45 &  32 &  $-$13.5 &  39.6 &      43~$\pm$~7 &                 \\
 2\dotfill &      10 27 50.6 &   $-$43 54 19.8 &                   $<0.38$ &       1.2$^{+0.7}_{-0.3}$ &   60 &                   $<0.20$ &       2.0$^{+2.1}_{-0.7}$ &  60 &  55 &  $-$13.2 &  39.9 &      67~$\pm$~8 &                 \\
 3\dotfill &      10 27 50.8 &   $-$43 54 15.2 &    0.37$^{+0.37}_{-0.32}$ &       2.2$^{+0.4}_{-0.3}$ &   44 &    0.08$^{+0.24}_{-0.08}$ &       1.2$^{+0.7}_{-0.3}$ &  46 &  52 &  $-$13.5 &  39.7 &      71~$\pm$~8 &                 \\
 4\dotfill &      10 27 51.2 &   $-$43 54 13.9 &    0.36$^{+0.14}_{-0.12}$ &       2.8$^{+0.4}_{-0.4}$ &  123 &    0.04$^{+0.09}_{-0.04}$ &       0.8$^{+0.1}_{-0.1}$ & 126 & 140 &  $-$13.1 &  40.1 &    329~$\pm$~18 &     Nucleus (N) \\
 5\dotfill &      10 27 51.6 &   $-$43 54 09.3 &    0.32$^{+0.18}_{-0.16}$ &       2.9$^{+0.6}_{-0.6}$ &   65 &    0.02$^{+0.11}_{-0.02}$ &       0.8$^{+0.2}_{-0.2}$ &  69 &  88 &  $-$13.5 &  39.6 &    139~$\pm$~12 &                 \\
 6\dotfill &      10 27 51.8 &   $-$43 54 13.3 &    0.55$^{+0.25}_{-0.21}$ &       3.6$^{+0.8}_{-0.8}$ &   60 &    0.18$^{+0.15}_{-0.13}$ &       0.6$^{+0.2}_{-0.1}$ &  63 &  68 &  $-$13.7 &  39.4 &    118~$\pm$~11 &                 \\
 7\dotfill &      10 27 55.1 &   $-$43 54 46.6 &                   $<0.71$ &       0.5$^{+1.4}_{-0.8}$ &   33 &                   $<0.43$ &                   $>1.35$ &  33 &  25 &  $-$13.1 &  40.0 &      33~$\pm$~6 &                 \\
\hline
 \multicolumn{14}{c}{NGC~3310}   \\
\hline
 1\dotfill &      10 38 43.3 &     +53 31 02.0 &    0.22$^{+0.15}_{-0.14}$ &       2.3$^{+0.4}_{-0.4}$ &   31 &    0.01$^{+0.10}_{-0.01}$ &       0.9$^{+0.2}_{-0.2}$ &  27 &  23 &  $-$12.8 &  39.8 &    168~$\pm$~13 &                 \\
 2\dotfill &      10 38 44.8 &     +53 30 04.3 &    0.55$^{+0.54}_{-0.49}$ &       1.6$^{+0.7}_{-0.7}$ &   12 &    0.34$^{+0.37}_{-0.16}$ &       1.7$^{+2.0}_{-0.6}$ &  11 &  10 &  $-$12.7 &  40.0 &    152~$\pm$~12 &                 \\
 3\dotfill &      10 38 45.8 &     +53 30 12.2 &    0.30$^{+0.71}_{-0.30}$ &       0.4$^{+0.7}_{-0.6}$ &    9 &    0.58$^{+0.35}_{-0.30}$ &                     $>10$ &  10 &   8 &  $-$12.4 &  40.3 &    174~$\pm$~13 &         Nucleus \\
 4\dotfill &      10 38 46.0 &     +53 30 04.8 &    0.71$^{+0.42}_{-0.38}$ &       1.8$^{+0.6}_{-0.6}$ &    9 &    0.40$^{+0.28}_{-0.25}$ &       1.5$^{+0.9}_{-0.4}$ &  10 &  16 &  $-$12.6 &  40.1 &    131~$\pm$~11 &                 \\
 5\dotfill &      10 38 46.5 &     +53 30 38.3 &    0.84$^{+0.51}_{-0.46}$ &       2.0$^{+0.7}_{-0.6}$ &   14 &    0.51$^{+0.35}_{-0.31}$ &       1.3$^{+0.7}_{-0.3}$ &  12 &  12 &  $-$12.7 &  40.0 &    142~$\pm$~12 &                 \\
 6\dotfill &      10 38 46.6 &     +53 30 13.7 &    0.59$^{+0.67}_{-0.59}$ &       1.7$^{+0.8}_{-0.8}$ &    2 &    0.24$^{+0.44}_{-0.24}$ &       1.9$^{+2.4}_{-0.7}$ &   3 &   8 &  $-$12.8 &  39.9 &    130~$\pm$~11 &                 \\
 7\dotfill &      10 38 46.7 &     +53 30 07.8 &                   $<0.22$ &       1.9$^{+0.7}_{-0.4}$ &    5 &                   $<0.08$ &       0.8$^{+0.3}_{-0.2}$ &   8 &   7 &  $-$13.1 &  39.5 &    213~$\pm$~15 &                 \\
 8\dotfill &      10 38 47.2 &     +53 30 27.9 &    0.28$^{+0.33}_{-0.28}$ &       1.8$^{+0.6}_{-0.6}$ &   13 &    0.04$^{+0.22}_{-0.04}$ &       1.5$^{+0.7}_{-0.4}$ &  13 &  10 &  $-$12.9 &  39.8 &    121~$\pm$~11 &                 \\
 9\dotfill &      10 38 50.2 &     +53 29 26.0 &    0.46$^{+0.34}_{-0.32}$ &       2.0$^{+0.7}_{-0.7}$ &   22 &    0.25$^{+0.24}_{-0.22}$ &       1.0$^{+0.5}_{-0.3}$ &  19 &  12 &  $-$12.9 &  39.8 &    294~$\pm$~17 &                 \\
\hline
\end{tabular}
\end{center}
NOTE.---All quoted errors on fit parameters indicate 90\% confidence intervals, while quoted errors associated with net counts are 1$\sigma$. Col.(1): Source ID. Col.(2) and (3): Right ascention and declination, respectively, based on {\it Chandra} source location. Col.(4)--(6) provide parameters to spectral fits for a power-law with intrinsic absorption model.  Col.(4): Intrinsic absorption column density in units of $10^{22}$~cm$^{-2}$.  Col.(5): power-law photon index. Col.(6): Minimum {\ttfamily cstatistic} $C$. Col.(7)--(9) provide parameters to spectral fits for a multicolor accretion disk ({\ttfamily diskbb}) with intrinsic absorption model. Col.(7): Intrinsic absorption column density in units of $10^{22}$~cm$^{-2}$.  Col.(8): Best-fitting inner accretion disk temperature ($kT_{\rm in}$) in units of keV. Col.(9): Minimum {\ttfamily cstatistic} $C$. Col.(10) Number of degrees of freedom for both power-law and multicolor accretion disk fits. Col.(11) and (12): Observed \hbox{2--10~keV} flux and luminosity based on the best-fitting model. Col.(13) Net counts in the 0.5--7~keV \chandra\ band. Col.(14) Notes on the individual sources.\\
\vspace{0.14in}
\end{table*}

\subsection{{\itshape Chandra} Point-Source Analyses}

Our joint \nustar\ and \chandra\ spectral analyses indicate that the
galaxy-wide 0.3--30~keV emission from NGC~3256 and NGC~3310 are dominated by
the combination of hot gas and ULX populations, with no obvious signatures of
luminous AGN.  In this section, we gain further insight into the \xray\
emitting populations in these galaxies by performing basic \chandra\ spectral
analyses of the detected point-sources and diffuse
emission.  

For each galaxy, we identified point sources that had $>$20 \hbox{0.5--7~keV}
counts that were also detected in the \hbox{2--7~keV} band images using
{\ttfamily wavdetect} at a false-positive probability threshold of $10^{-5}$;
these sources were candidates for performing basic spectral fits.  For NGC~3256
and NGC~3310, we found 7 and 9 sources, respectively, that satisfied these
criteria; the locations of these sources are annotated in Figure~3.  Taken
together, these sources respectively provide $\approx$42\% and $\approx$79\% of
the \hbox{2--7~keV} \chandra\ net counts within the total galaxy apertures
defined in $\S$4.1.  For these sources, we extracted the \chandra\ spectra and
fit them using both an absorbed power-law model ({\ttfamily powerlaw}) and an
absorbed multicolor accretion disk ({\ttfamily diskbb}; Shakura \&
Sunyaev~1973; Mitsuda \etal\ 1984).  In this procedure, point-sources were
extracted from circular regions that encompassed 90\% of the encircled energy
of the PSF ($\approx$2~arcsec in radius for all sources).  We utilized the same
background regions used in $\S$4.1 to estimate background spectra for the
sources.  As such, some source regions will contain background related to
diffuse emission.  Since the exposures are relatively shallow and diffuse
emission gradients are high, we chose not to attempt to model the local
background component.  When relevant, we have qualified interpretations of the
\chandra\ source spectra with this limitation.  Given the small number of
counts per source (33--329 0.5--7~keV counts), all spectral fits were performed
by minimizing the Cash statistic ({\ttfamily cstat}; Cash~1979) using spectra binned to a minimum of 1~count per channel.

In Table~2, we present the best-fit parameters for the point-source spectral
fits.  We note that the parameters in Table~2 provide only a basic description
of the data and are not well constrained, e.g., due to degeneracies between
column density and spectral slope or inner disk temperature.  In particular,
for sources that were located near the nuclei of the galaxies, some
contributions from hot gas are expected, which will steepen the inferred
spectral slopes.  In the majority of cases, a simple power-law provided an
acceptable fit to the spectra; we find a slight ($\approx$1~$\sigma$ level)
statistical preference for the {\ttfamily diskbb} model for sources 1 and
9 of NGC~3310.  The luminosity range of the point-source populations cover
$L_{\rm 2-10~keV} =$~(3--20)~$\times 10^{39}$~\lum, indicating that if all
sources are discrete objects, then they are all in the ULX range of
luminosities.  The photon indices for the point sources range from $\Gamma
=$~\hbox{0.4--3.6} (median $\langle \Gamma \rangle = 1.8$), with only source~3
in NGC~3310 having $\Gamma < 1.5$ at $>$90\% confidence.  As we will discuss in
the next section, source~3 is coincident with the nuclear region in NGC~3310
and is a candidate obscured AGN.  


\subsection{Contributions from Nuclear Sources and Potential AGN}

Both NGC~3256 and NGC~3310 contain \chandra\ detected point-like sources
coincident with their galactic nuclei (NGC~3256 source~4 and NGC~3310 source~3); however,
for NGC~3256, we only detect the northern nucleus and not the southern nucleus.
In this section, we investigate the nature of the sources coincident with the nuclei using \chandra\
spectral constraints and discuss their potential contributions to the 
\hbox{0.3--30~keV} spectra.

For NGC~3256, there is a clear detection of the northern nucleus, which is the
brightest point-like source in our \chandra\ catalog (source~4).  Using a
deeper $\approx$30~ks \chandra\ exposure (ObsID~835), Lira \etal\ (2002) found
that the nuclear emission is clearly extended, suggesting that the emission
must include extended and/or multiple point-sources.  From Table~2, we see the
\chandra\ spectrum of this source is fit well with a steep power-law ($\Gamma
= 2.8 \pm 0.4$) with modest absorption ($N_{\rm H} \approx 4 \times
10^{21}$~cm$^{-2}$).  Such a steep photon index is uncharacteristc of AGN,
which typically span $\Gamma \approx$~1--2.2.  We refit the spectrum using only
the 3--7~keV range, to mitigate any biases in the spectral fit due to soft
emission from the hot gas, which will inevitably be coincident with our
spectral extraction of this region.  The resulting best-fit photon index
remained unchanged, indicating that there is no evidence for an emerging hard
\xray\ AGN in this region.  This result is consistent with previous studies
that have concluded that the northern nucleus in NGC~3256 is powered almost
exclusively by star-formation activity (e.g., Lira \etal\ 2002; Jenkins \etal\
2004; Sakamoto \etal\ 2014).  In Figure~4, we show the contribution from the
northern nucleus extrapolated through to the \nustar\ band.  Based on this
extrapolation and the global fit presented in $\S$4.1, we estimate that this
source will contribute $\approx$4--10\% of the galaxy-wide emission across the
full 0.3--30~keV band.

Our 16~ks \chandra\ observation of NGC~3256 did not yield a clear detection of
the southern nucleus (location indicated in the left panel of Fig.~3), which
has previously been shown to have submm and near-infrared evidence for a
collimated outflow, potentially due to an AGN that is currently inactive (e.g.,
Sakamoto \etal\ 2014; Emonts \etal\ 2014).  Two previous deeper $\approx$25~ks
archival \chandra\ exposures (ObsIDs 835 and 3569) reveal clear detections of a
source coincident with the southern nucleus, suggesting that the source does
produce powerful, albeit heavily obscured, \xray\ emission at the
$\approx$$10^{40}$~\lum\ level (in the 0.3--10~keV band).  Using the \chandra\
exposure from ObsID 835, as well as radio, near-IR, and optical data, Lira
\etal\ (2002) showed that this source is likely to be a low luminosity AGN that
does not provide a significant contribution to the total \xray\ emission of the
galaxy.  No evidence was found implicating a luminous, heavily obscured AGN
(e.g., Fe-line emission).  The non-detection of the southern nucleus in our
16~ks \chandra\ exposure indicates a change in the southern source with respect
to the previous two observations.  In 2000~Jan~5 and 2003~May~23, the southern
source was detected with similar 2--7~keV \chandra\ count-rates of $(2.7 \pm
0.3) \times 10^{-3}$~cnts~s$^{-1}$ and $(2.9 \pm 0.3) \times
10^{-3}$~cnts~s$^{-1}$, respectively.  Whereas the 3$\sigma$ upper limit on the
2--7~keV \chandra\ count-rate in our 2014~Aug~28 observation is $<$$1.2 \times
10^{-3}$~cnts~s$^{-1}$, indicating a factor of $\simgt$2.0--2.5 times decline
in the southern source hard-band intensity.  As shown by our \nustar\ spectrum,
the steep spectral slope of the full galaxy persists out to at least
$\approx$30~keV, limiting the possible contributions from a hard southern
nuclear AGN.  This implies that if there is an AGN associated with the southern
nucleus, it is likely to be of low luminosity.

To place more stringent constraints on the spectral properties of the southern
nucleus, we combined the previous two \chandra\ observations of NGC~3256 to
create a merged $\approx$50~ks data set.  
We extracted the merged spectrum of
the southern nuclear source and performed spectral fitting in {\ttfamily
XSPEC}.  The spectrum can be fit well by a simple power-law model with an
inverted photon index of $\Gamma \approx -0.1$; the best-fit model implies a
2--10~keV luminosity of $L_{\rm 2-10~keV} \approx 10^{40}$~\lum.  It is likely
that the spectrum contains both contributions from hot gas and a
heavily obscured power-law source coincident with the nuclear region.  The
black hole mass for the southern nucleus is on the order of
$\sim$$10^{7}$~$M_\odot$ (e.g., Alonso-Herrero \etal\ 2013), implying an
Eddington luminosity of $\sim$$10^{45}$~\lum.  If the southern nucleus is
indeed an AGN it would therefore have $L_{\rm 2-10~keV}/L_{\rm Edd} \sim
10^{-5}$, and the expected intrinsic spectral slope of this source would be
between $\Gamma =$~1.5--2.0 (Younes \etal\ 2011).  
Using these constraints, we
fit the spectrum of the southern nucleus using an absorbed plasma plus an
absorbed power-law model ({\ttfamily tbabs*apec+ tbabs*powerlaw} in {\ttfamily
XSPEC}).  We held the plasma temperature fixed at 0.6~keV (see Lira \etal\ 2002
for motivation) and the power-law slope fixed at $\Gamma = 1.8$ and fit for the
component normalizations and obscuring columns.  This model yielded a good fit
to the data for column densities $N_{\rm H, gas} \approx 10^{22}$~cm$^{-2}$ and
$N_{\rm H, PL} \approx 5 \times 10^{22}$~cm$^{-2}$.  

In the left panel of Figure~4, we show the \chandra\ spectrum and best-fitting
model of the southern nucleus of NGC~3256 from the previous two observations
when the source was detectable ({\it orange data and curve\/}).  The data and
model have been scaled down by a factor of 2.5 to show the upper limit to our
2014 observation (see above).  Based on an extrapolation of this model to the
\nustar\ band, the southern nucleus is expected to contribute $\simlt$10--30\%
of the \nustar\ emission at 10--30~keV.  Such a low level of AGN activity is
insufficient to have any material effect on the overall shape of the galaxy
spectrum nor on the constraints presented in $\S$4.1 and Table~1.  We therefore
expect that the 0.3--30~keV spectrum of NGC~3256 is dominated by non-AGN
emitting sources (i.e., hot gas and \xray\ binaries).

For NGC~3310, a previous $\approx$50~ks \chandra\ observation from 2003
(ObsID~2939) was used by Tzanavaris \& Georgantopoulos~(2007) to show that
there is evidence for an Fe-K$\alpha$ emission line in the nuclear source
(source~3), implying the presence of a hidden AGN.  Our $\approx$10~ks
\chandra\ observation was too shallow to verify the claimed Fe-K$\alpha$
feature, which was of low significance ($\approx$2~$\sigma$) in the much deeper
2003 observation.  However, we do find that the best-fitting power-law spectral
slope of the nuclear point-source $\Gamma \approx 0.4$ is much shallower than
the other sources in the galaxy ($\Gamma \approx$~1.6--2.3), and the luminosity
of the source ($L_{\rm 2-10~keV} \approx 10^{40}$~\lum) is the highest of all
the point sources.  These facts are consistent with an obscured or low
luminosity AGN hypothesis.  We note that if we invoke a more complex model
appropriate for AGN that includes both direct and scattered components (e.g., {\ttfamily
pexrav} or {\ttfamily MyTorus}; Magdziarz \& Zdziarski~1995; Murphy \&
Yaqoob~2009), the expected contribution to the galaxy-wide emission above
10~keV can quickly approach 100\%, and the intrinsic emission could be much
higher than the observed luminosity inferred by \chandra.

%
%
\begin{figure}
\figurenum{5}
\centerline{
\includegraphics[width=8cm]{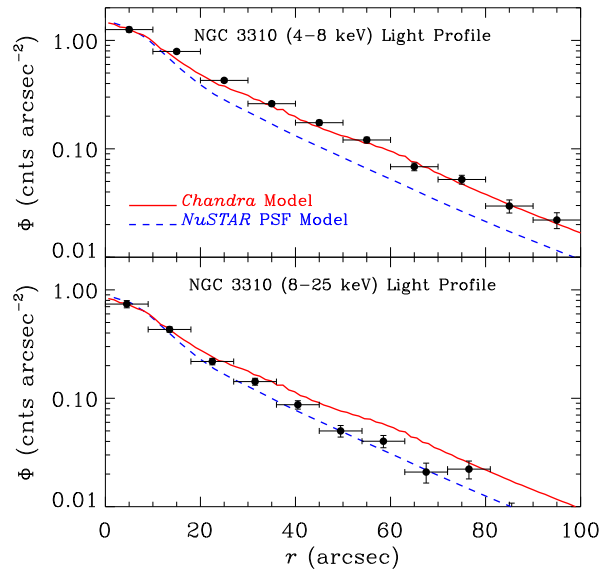}
}
\caption{
\nustar\ light profiles for NGC~3310 in the 4--8~keV ({\it top\/}) and
8--25~keV ({\it bottom\/}) bands.  A blurred PSF model ({\it blue dashed
curves\/}) and a model based on the relative \hbox{4--8~keV} \chandra\ point-source
brightnesses (``Chandra Model''; {\it red solid curves\/}) have been compared
with the \nustar\ light profiles ({\it black points\/}).  We find that the ``Chandra Model'' characterizes
well the \nustar\ galaxy light profile in the \chandra\ and \nustar\
overlapping 4--8~keV bandpass.  However, due to the presence of harder sources
in the central region of the galaxy, as compared with sources in the outskirts,
the 8--25~keV light profile becomes more point-like.  Such a result could be
explained by the presence of an obscured AGN and/or harder ULXs in the central
region of the galaxy.
}
\end{figure}

Following a similar approach to that taken for the southern nucleus in
NGC~3256, we tested the influence that the nuclear source in NGC~3310 would
have on the $\approx$7--30~keV emission, if it were truly an AGN, by
extrapolating the \chandra\ spectrum through to the \nustar\ band.  The black
hole mass of NGC~3310 is currently poorly constrained, but has been estimated
via gas kinematics to be $\sim$(5--40)~$\times 10^6$~\msol, which implies an
Eddington luminosity of $\sim$(6--50)~$\times 10^{45}$~\lum\ (Pastorini \etal\
2007).  The 2--10~keV luminosity of the source, as measured by \chandra, is
$L_{\rm 2-10~keV} \sim 10^{40}$~\lum, implying $L_{\rm
2-10~keV}/L_{\rm Edd} \sim 10^{-6}$--$10^{-5}$.  From Younes \etal\ (2011),
this implies an intrinsic photon index of $\Gamma =$~\hbox{1.5--2.2}.  Following the
approach used for the southern nucleus of NGC~3256, we fit the \chandra\
spectrum of the NGC~3310 nuclear source using an absorbed plasma plus an
absorbed power-law model ({\ttfamily tbabs*apec+ tbabs*powerlaw} in {\ttfamily
XSPEC}) with the plasma temperature fixed at 0.3~keV and the power-law slope
fixed at $\Gamma = 1.8$.  These parameters yielded a good fit to the data for
column densities $N_{\rm H, gas} \approx 5 \times 10^{21}$~cm$^{-2}$ and
$N_{\rm H, PL} \approx 3 \times 10^{22}$~cm$^{-2}$.

In the right panel of Figure~4, we show the \chandra\ spectrum for the nuclear
source of NGC~3310 and the best-fit model extrapolated across the full
0.3--30~keV band.  It is clear that the nuclear source is expected to provide a
non-negligible contribution to the \xray\ emission above \hbox{4--5~keV}.
Based on our extrapolated model, the point-source makes a fractional
contribution of $\simlt$3\% of the galaxy-wide emission at \hbox{$E < 2$~keV} and
reaches a maximum contribution of \hbox{$\approx$30--100\%} in the \hbox{20--30~keV} band.  If we
revist our fits to the galaxy-wide spectrum performed in $\S$4.1 and now fix
the diffuse component and best-ftting model for the nuclear source, we find
that the remaining spectrum associated with the bright point sources can be
well characterized by a broken power-law with parameters consistent with those
presented in Table~1.

\subsection{Spatially Extended \nustar\ Emission for NGC~3310}

%
%
\begin{figure*}
\figurenum{6}
\centerline{
\includegraphics[width=18cm]{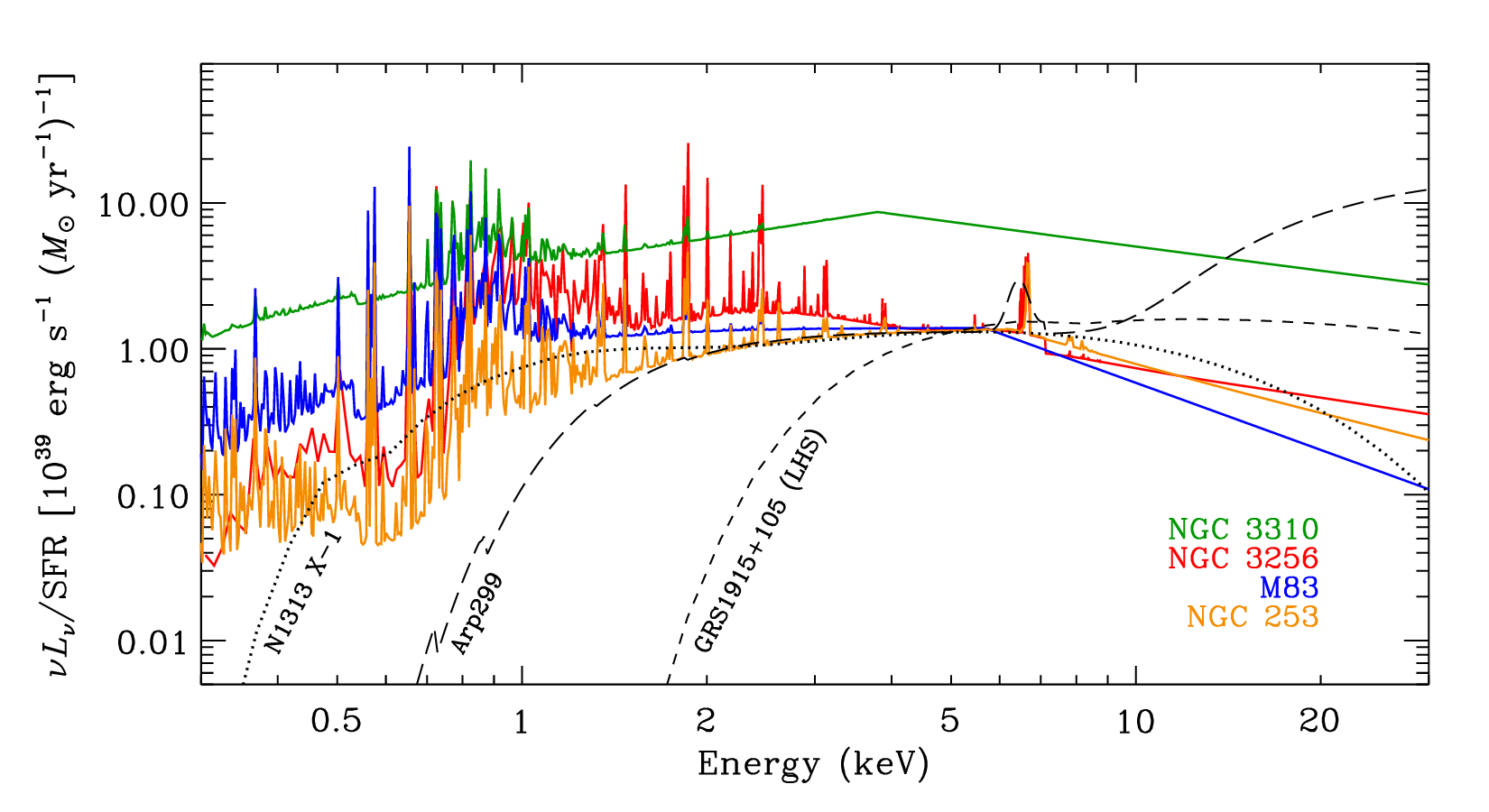}
}
\caption{
SFR-normalized 0.3--30~keV SEDs for NGC~3256 ({\it red\/}), NGC~3310 ({\it
green\/}), NGC~253 ({\it orange\/}), and M83 ({\it blue\/}).  For illustrative
purposes, we have displayed only the best-fit SEDs to the nearly simultaneous
\chandra/\xmm\ plus \nustar\ data.  NGC~253 and M83 fits are taken from the
best-fit models presented in Wik \etal\ (2014a) and Yukita \etal\
(in-preparation).  For comparison, we have shown \nustar\ constrained SEDs for
the ULX NGC~1313 X-1 ({\it dotted curve}; Bachetti \etal\ 2013), the luminous
black-hole \xray\ binary GRS~1915+105 in the low/hard state ({\it short-dashed
curve}; Miller \etal\ 2013), and the Compton-thick AGN in Arp~299 ({\it
long-dashed curve}; Ptak \etal\ 2014); these SEDs have been normalized to $\nu
L_\nu$/SFR~=~$1.3 \times 10^{39}$~\lum~($M_\odot$~yr$^{-1}$)$^{-1}$ at 5~keV
for ease of comparison.  In this representation, a power-law SED with
photon-index $\Gamma = 2$ is a flat line.  It is clear that all four galaxies
here show steepening in their SEDs above $\approx$3--6~keV, where their
spectral slopes reach $\Gamma \approx$~2.5--3.0 due primarily to the ULX
population.  These constraints indicate that the ULX population on average has
a similar spectral shape to that of super-Eddington accreting sources like
NGC~1313 X-1.
}
\end{figure*}

Although the \nustar\ PSF is too large to identify individual \xray\ point
sources in the galaxies studied here, it is sufficient for identifying extended
features due to the $\approx$50~arcsec wide distribution of point sources throughout NGC~3310.
To test for extended \nustar\ emission, we first
constructed a spatial model of the expected \nustar\ emission profile within a
bandpass that overlapped with \chandra\ (i.e., \hbox{4--8~keV}), and then tested
to see how the \nustar\ emission profile changed in a bandpass above the
\chandra\ response (i.e., \hbox{8--25~keV}).  If the \xray\ SEDs of the underlying point
sources were all similar, we would expect the profile to remain the same for the
soft and hard bands.  Using the positions and relative 4--8~keV \chandra\
count-rates for the point sources, we constructed a model distribution map of
\nustar\ counts.  This map was constructed by co-adding \nustar\ PSF images centered at
the locations of the \chandra\ point sources, with normalizations scaled
by the 4--8~keV
\chandra\ count-rates.  Throughout this process, our \nustar\ PSFs were
constructed by accounting for the off-axis angles of the sources and include
errors associated with variations in off-axis angle
due to the
motions of the mast.  We note that some relative astrometric offset is expected
to be present between the \nustar\ and \chandra\ data.  In order to estimate
this error, we fit Gaussian profiles to both the \nustar\ 4--8~keV image
emission and the \chandra-based model distribution map to identify the location
of the peak emission.  We then extracted the emission profiles for the images
and models and compared them.

In Figure~5, we show the spatial distribution of \nustar\ counts for the
\hbox{4--8~keV} and \hbox{8--25~keV} bands.  The expected distributions for the
\chandra-based model (described above) and an individual point-source PSF model
have been overplotted for comparison.  By construction, the 4--8~keV \nustar\
emission profile follows the \chandra-based model well and is statistically
inconsistent with a single point source.  However, at 8--25~keV, the \nustar\
emission profile is more consistent with a single point-source (i.e., the
\nustar\ PSF) than with the \chandra-based model, implying that the central
region of NGC~3310 contains a harder \xray\ spectrum (across 8--25~keV)
compared to point-sources in the outskirts of the galaxy.  Our spectral
extrapolation of the nuclear source presented in $\S$4.3 and Figure~4 ({\it
right\/}) shows that the fractional contribution that the source makes to the
galaxy-wide emission increases with energy.  It is also the case that the two
sources in the outskirts of the galaxy (sources~1 and 9; see Fig.~3 and
Table~2) have the steepest spectral slopes out of all nine point sources.
Therefore, the hard spectra of the nuclear source and the five relatively hard
non-nuclear sources that are within $\approx$20~arcsec of the galactic center
are likely to be responsible for the transition to a more point-like emission
profile with increasing energy.  As we discussed above, we find a steep $\Gamma
\approx 2.7$ power-law slope at $E \simgt 4$~keV for the galaxy-wide spectrum
of NGC~3310, which is steeper than the best fit single power-law slopes
($\Gamma \approx$~0.3--2.0) derived for all five sources.  This implies that,
although the spectra of these sources may be intrinsically hard in the
\chandra\ band, spectral turnovers at $E \simgt$3--10~keV must be typical of
these sources, making it difficult to quantify the relative influences these
sources have on the emission measured in the \nustar\ band.

%
\section{Discussion}
%

We have utilized nearly simultaneous \chandra\ and \nustar\ observations of the
powerful local starburst galaxies NGC~3256 and NGC~3310 to investigate the
nature of the \xray-emitting components across the broad \hbox{0.3--30~keV}
bandpass.  Similar to previous \xmm\ and \chandra\ studies, we find significant
$kT \approx$~0.2--1~keV plasma emission that dominates the spectra of these
galaxies out to $\approx$3~keV and $\approx$1~keV for NGC~3256 and NGC~3310,
respectively.  At the higher energies probed by \nustar\ ($E \simgt 3$~keV), we
find the majority of the emission is produced by populations of the brightest
\hbox{5--10} ULXs.  The cumulative spectra of both galaxies exhibit steep
power-law slopes ($\Gamma \approx$~2.6--2.7) above $\approx$3--6~keV that are
similar to those found in \nustar\ studies of individual ULXs, implying that no
strong hard component exists for either galaxy.  This places stringent limits
on the influence of possible AGN candidates in both galaxies, which are
expected to have much flatter spectral slopes in this band.  Sensible
extrapolations of the \chandra\ spectra of AGN candidates in these galaxies,
the southern nucleus of NGC~3256 and the nucleus of NGC~3310, indicate that
these candidates provide only minority contributions to
the galaxy-wide \nustar\ emission out to 30~keV.  Regardless of extrapolation
assumptions, our constraints indicate that the supermassive black holes in
these galaxies are of low intrinsic luminosity and are accreting at the
$\simlt$$10^{-6}$--$10^{-5}$ Eddington levels.  Our spectral constraints on
NGC~3256 and NGC~3310 therefore provide first measurements of the 0.3--30~keV
spectra of the non-AGN \xray\ emitting populations within powerful starburst
galaxies with high-sSFRs.

In Figure~6, we show the SFR-normalized best-fitting \hbox{0.3--30~keV} SEDs
(i.e., $\nu L_\nu$/SFR) of NGC~3256 and NGC~3310.  For comparison, we also
provide best-fit models for M83 (Yukita \etal\ in-preparation) and NGC~253 (Wik
\etal\ 2014a), which were also constrained using nearly simultaneous
observations with \chandra/\xmm\ and \nustar.  The displayed SEDs have not been
corrected for intrinsic absorption and are therefore representative of the {\it
observed} \xray\ spectra.  In the low-energy regime ($E \simlt$~2--3~keV), all
four galaxies have significant contributions from line-emitting hot plasmas.
With the exception of NGC~253 ({\it orange curve\/}) the SFR-normalized gas
components have similar normalizations around $\approx$1~keV, consistent with
previous studies that have found a direct scaling of the hot gas emission with
SFR (see, e.g., Mineo \etal\ 2012b).  Compared with the other three galaxies,
NGC~253 is relatively edge-on ($i \approx 80$~deg), and therefore absorption is
likely playing a role in the apparent deficit of hot gas emission in the
galaxy.  

At higher energies, $E \approx$~3--30~keV, we find that all galaxies, except
for NGC~3310 (see discussion below), have SFR-normalized SEDs that are in good
agreement with each other.  For all four galaxies, the steepening of the
spectral slope in the $\approx$6--30~keV range is clearly apparent.  For M83,
it appears that the spectral slope in this regime is somewhat steeper than it
is for the other three galaxies.  As discussed by Yukita \etal\
(in-preparation), the spectral shape of M83 is strongly influenced by a single
variable ULX (see also Soria \etal\ 2012 for details on this source).  For
comparison, we have overlaid in Figure~6 the \nustar-constrained SEDs (scaled
to the mean starburst galaxy emission at 5~keV) of the ULX \hbox{NGC~1313~X-1}
(Bachetti \etal\ 2013), the luminous black-hole X-ray binary GRS 1915+105 in
the low/hard state (Miller \etal\ 2013), and the Compton-thick AGN in Arp~299
(Ptak \etal\ 2014).  From this illustration, it is clear that the $>$10~keV
starburst galaxy SEDs, including the turnover in spectral slope, are very
similar to the SED of NGC~1313 X-1, a likely super-Eddington accreting ULX.
Given the fact that ULXs clearly dominate the point source emission from this
galaxy population, our observations constrain the average spectral shape for
the ULX population, in general.

Interestingly, for NGC~3310, we observe a clear factor of $\approx$3--10 times excess \xray\
emission per unit SFR over the $\approx$6--30~keV range.  The excess is the
result of an overabundance of ULXs in the galaxy compared to typical galaxies.
From Mineo \etal\ (2012a), we expect that there would be $\approx$1--4 ULXs
given the SFR of NGC~3310 (SFR~$\approx$~6~\sfr); however, 9 ULXs are clearly
detected, indicated a significant excess in both the \xray\ emission and the
number of ULXs.  In general, an excess of luminous \xray\ binaries can be
explained by either (1) a star-formation history that is heavily weighted
towards an epoch when ULXs are expected to be most luminous due to the presence
of the most massive donor stars ($\approx$5--10~Myr) and/or (2) star-formation
that is happening in low-metallicity environments, in which an excess
population of black holes form (e.g., Linden \etal\ 2010).  

\hst-based studies of the young star cluster properties in NGC~3310, using SED
fitting of UV--to--near-IR \hst\ data, indicate a peak intensity in the recent
star-formation history around $\approx$30~Myr ago (de~Grijs \etal\ 2003a,b),
well past the peak of ULX activity.  Furthermore, similar star-cluster ages are
also seen in NGC~3256 (e.g., Trancho \etal\ 2007), which suggests that the
excess of luminous \xray\ binaries in NGC~3310 is unlikely to be caused by
differences in recent star-formation history alone.  By contrast, the
metallicity distribution of young star clusters in NGC~3310 has been estimated
to peak at $Z \approx 0.4$~$Z_\odot$ (de~Grijs \etal\ 2003a,b), compared with
$Z \approx 1.5$~$Z_\odot$ for M83, NGC~253, and NGC~3256 (e.g., Zaritaas \etal\
1994; Boselli \etal\ 2002; Bresolin \etal\ 2014; Trancho \etal\ 2007).  The
relatively low metallicity in NGC~3310 is the result of the recent
cannibalization of a low-metallicity dwarf galaxy that triggered the current
star-formation event.  It is therefore plausible that the lower metallicity of
NGC~3310 is the driving mechanism behind the excess of ULXs per unit SFR in
this galaxy.  Indeed, past studies of both the number of ULXs and galaxy-wide
\xray\ luminosity per unit SFR have revealed excesses at low metallicities
(e.g., Mapelli \etal\ 2009; Prestwich \etal\ 2013; Basu-Zych \etal\ 2013b;
Brorby \etal\ 2014).  From the population synthesis models from Fragos \etal\
(2013b), galaxies with $Z \approx 1.5$~$Z_\odot$ and $0.4$~$Z_\odot$ are
predicted to have SFR-normalized \hbox{2--10~keV} luminosities of $L_{\rm
2-10~keV}$/SFR~$=$~\hbox{(1.1--3.4)}~$\times 10^{39}$~\lum\ and
\hbox{(3.5--11.2)~$\times 10^{39}$~\lum}, respectively (e.g., see their
Fig.~2).  We measure corresponding values of $L_{\rm 2-10~keV}$/SFR~$=$~$2.1
\times 10^{39}$~\lum\ and $7.3 \times 10^{39}$~\lum, respectively, in very good
agreement with the theoretical model predictions.  We therefore favor
metallicity as being the dominant underlying factor responsible for the
observed \hbox{$\approx$3--10} times excess \xray\ emission per unit SFR in
NGC~3310; however, 
future investigations of the properties of the immediate environments (i.e.,
characteristic stellar ages and metallicities) near the ULXs would help to
better discriminate the roles that age and metallicity play in producing the
excess ULX population.

\acknowledgements

We thank the anonymous referee for helpful comments, which have improved
the quality of this paper.  We gratefully acknowledge financial support from
\chandra\ X-ray Center grant GO4-15086Z (B.D.L., J.B.T.) and NASA ADAP grant
NNX13AI48G (B.D.L.).  A.Z.  acknowledges funding from the European Research
Council under the European Union's Seventh Framework Programme
(FP/2007-2013)/ERC Grant Agreement n.~617001.  This work was supported under
NASA Contract No.  NNG08FD60C, and made use of data from the {\it NuSTAR}
mission, a project led by the California Institute of Technology, managed by
the Jet Propulsion Laboratory, and funded by the National Aeronautics and Space
Administration.  This research has made use of the \nustar\ Data Analysis
Software (NuSTARDAS) jointly developed by the ASI Science Data Center (Italy)
and the California Institute of Technology (USA).

{\it Facilities:} Chandra, NuSTAR

%

%

\end{document}